\documentclass[11pt, a4paper]{article}

\usepackage[a4paper, total={16cm, 25cm}]{geometry}
\usepackage{amsmath}
\usepackage{graphicx}
\usepackage{amsthm}
\usepackage{color}
\usepackage{footmisc}

\newtheorem{theorem}{Theorem}

\newtheorem{rational for conjecture}{Rational for Conjecture}

\begin{document}

\LARGE
\begin{center}
  \textbf{Quantile Tracking in Dynamically Varying Data Streams Using a Generalized Exponentially Weighted Average of Observations} \\[10mm]
\end{center}

\large

\begin{center}
  Hugo Lewi Hammer\footnote[2]{OsloMet -- Oslo Metropolitan University}\footnote[4]{Corresponding author. Email: \texttt{hugo.hammer@oslomet.no}}, Anis Yazidi\footnotemark[2] and H{\aa}vard Rue\footnote[5]{King Abdullah University of Science \& Technology}
\end{center}

\normalsize

\begin{abstract}

The Exponentially Weighted Average (EWA) of observations is known to be state-of-art estimator for tracking \emph{expectations} of dynamically varying data stream distributions. However, how to devise an EWA estimator to rather track \textit{quantiles} of data stream distributions is not obvious. In this paper, we present a lightweight quantile estimator using a generalized form of the EWA. To the best of our knowledge, this work represents the first reported quantile estimator of this form in the literature. An appealing property of the estimator is that the update step size is adjusted online \textit{proportionally} to the difference between current observation and the current quantile estimate. Thus, if the estimator is off-track compared to the data stream, large steps will be taken to promptly get the estimator back on-track. The convergence of the estimator to the true quantile is proven using the theory of stochastic learning.

Extensive experimental results using both synthetic and real-life data show that our estimator clearly outperforms legacy state-of-the-art quantile tracking estimators and achieves faster adaptivity in dynamic environments. The quantile estimator was further tested on real-life data where the objective is efficient online control of indoor climate. We show that the estimator can be incorporated into a concept drift detector for efficiently decide when a machine learning model used to predict future indoor temperature should be retrained/updated.
\end{abstract}

\section{Introduction}

In this paper we consider the problem of tracking quantiles when data arrive sequentially (data stream). The problem has been considered for many applications like portfolio risk measurement in the stock market \cite{gilli2006application, abbasi2013bootstrap}, fraud detection \cite{zhang2008detecting}, signal processing and filtering \cite{stahl2000quantile}, climate change monitoring \cite{zhang2011indices}, SLA violation monitoring \cite{sommers2007accurate,sommers2010multiobjective}
and  back-bone network  monitoring \cite{choi2007quantile}.

Suppose that we are interested in estimating the quantile related to some probability $q$.
The most natural estimator is to use the $q$ quantile of the sample distribution.
Unfortunately, such a quantile estimator has clear disadvantages for data streams as computation time and memory requirement are linear to the number of samples received so far from the data stream. Such methods thus are infeasible for large data streams.

Several algorithms have been proposed to deal with those challenges. Most of the proposed methods fall under to the category of what can be called histogram or batch based methods. The methods are based on efficiently maintaining a histogram estimate of the data stream distribution such that only a small storage footprint is required. A thorough review of state-of-the-art histogram and batch methods is given in the related work section (Section \ref{sec:relwork}).

Another ally of methods are the so-called incremental update methods. The latter methods are based on performing small updates of the quantile estimate every time a new sample is received from the data stream. {\color{black} Generally,  the current estimate is a convex combination of the estimate at the previous time step and a quantity depending on the current observation.} One of the first and prominent examples of this family of methods is the algorithm attributed to Tierney (1983) \cite{Tierney1983} which is based on the stochastic learning theory. A few modifications of the Tierney method have been suggested, see e.g. \cite{Chen2000, cao2010tracking, cao2009incremental, Chambers2006}.

In data stream applications, a common situation is that the distribution of the samples from the data stream varies with time. Such system or environment is referred to as a dynamical system in the literature. Given a dynamical system, two main problems are considered in the literature namely to i) dynamically update estimates of quantiles of all data received from the stream so far or ii) estimate quantiles of the current distribution of the data stream (tracking). Despite the importance of efficient tracking of statistical properties, the tracking problem ii) has been far less studied in the literature than problem i). Incremental methods are well suited to address the tracking problem ii) while histogram and batch methods mainly have been used to address problem i). Histogram and batch based methods are not well suited for the tracking problem ii) and incremental methods typically are the only viable lightweight alternatives \cite{cao2009incremental}.

%Figure \ref{fig:1} shows an example to illustrate the tracking problem for dynamic data streams.
%\begin{figure}[h]
%  \centering
%  \includegraphics[width = \textwidth]{RealDataPlotUnivar}
%  \caption{The gray circles show the number of tweets posted by Norwegian Twitter users every minute from July 21 2011 to July 24 2011. The black, blue and red curves show running estimates of the 20, 50 and 80\% quantiles %of the distribution of the number of tweets posted.}
%  \label{fig:1}
%\end{figure}
%The gray circles show the number of tweets posted by Norwegian Twitter users every minute in the time period before and after the Oslo bombing and Ut{\o}ya massacre in Norway July 22 2011. The terror attack started by a bomb going off in Oslo at July 22 3:25 p.m, and, as expected, this was followed up by a rapid increase in the number of posted tweets. The black, blue and red curves show the tracking of the 20, 50 and 80\% quantiles of the distribution of the number of tweets posted using the method that will be presented in this paper. We see that the method efficiently tracks the quantiles of the distribution, even when the distribution changes rapidly.

Motivated by the lack of research on the tracking problem ii), the authors of this paper introduced the deterministic multiplicative incremental quantile estimator (DUMIQE) \cite{yazidi17multiplicative} given by
\begin{align}
\label{eq:1}
  \begin{split}
   \widetilde{Q}_{n+1}(q) &\leftarrow \widetilde{Q}_{n}(q) + \lambda q \widetilde{Q}_{n}(q)  \hspace{16mm}\text{ if } x_n > \widetilde{Q_{n}}(q) \\
   \widetilde{Q}_{n+1}(q) &\leftarrow \widetilde{Q}_{n}(q) - \lambda (1-q) \widetilde{Q}_{n}(q)  \hspace{6mm}\text{ if } x_n \leq \widetilde{Q_{n}}(q)
  \end{split}
\end{align}
The intuition behind the estimator is simple. If the received sample has a value below the current quantile estimate, the estimate is decreased. Alternatively, whenever the received sample has a value above the current quantile estimate, the estimate is increased. The ``weights'' $q$ and $1-q$ are included to ensure convergence to the true quantile. Even though the estimator is really simple, it can document state-of-the-art tracking performance \cite{yazidi17multiplicative}. However, as Eq. \eqref{eq:1} reveals, the estimator do not use the values of the received samples directly to update the estimate, but only whether the value of the samples are above or below some varying threshold. Intuitively, this seems like a waste of information received from the data stream. In this paper, we thus present an estimator that uses the values of the received samples directly. The estimator is such that the update step size is \textit{proportional} to the distance between the current estimate and the value of the sample. Thus if the current estimate is off-track compared to the data stream, the estimator will perform large jumps to rapidly get back on-track. A theoretical proof is provided to document the convergence properties of the estimator in addition to extensive simulation experiments. The experiments show that the estimator outperforms DUMQUE and several other legacy state-of-the-art quantile estimators.

The EWA of observations is known to be state-of-the-art estimator to track expectations of dynamically varying data streams \cite{gardner2006exponential}. Interestingly, we will show that the suggested quantile estimator in this paper is in fact an instance of a  \textit{generalized} EWA  such that \textit{quantiles} and \textit{not} expectations are tracked. To the best of our knowledge, this is the first EWA based quantile estimator found in the literature.

The paper is organized as follows. In Section \ref{sec:ewqe} we present the novel quantile estimator using an EWA of observations. In Section \ref{sec:qea}, we present a quantile estimation algorithms based on the estimator in Section \ref{sec:ewqe}. In Section \ref{sec:se}, we perform extensive experiments that document the superiority of the suggested algorithm. Finally, in Section \ref{sec:real-life} we apply the quantile estimator on real-life data related to the problem of efficient online control of indoor climate. More specifically the estimator is used to detect when a machine learning model should be retrained/updated which is commonly referred to as concept drift detection \cite{gama2014survey}.

\section{Related Work}
\label{sec:relwork}

In this Section, we shall review some of the related work on estimating quantiles from data streams. However, as we will explain later, these related works require some memory restrictions which renders our work radically distinct from them. In fact, our approach requires storing only one sample value in order to update the estimate.
The most representative work for this type of ``streaming'' quantile estimator is due to the seminal work of Munro and Paterson \cite{Munro1980}.
In \cite{Munro1980}, Munro and Paterson described a $p$-pass algorithm for selection using $O (n^{1/(2p)})$ space for any $p \ge 2$. Cormode and Muthukrishnan \cite{Cormode2005} proposed a more space-efficient data structure, called the Count-Min sketch, which is inspired by Bloom filters, where one estimates the quantiles of a stream as the quantiles of a random sample of the input. The key idea is to maintain a random sample of an appropriate size to estimate the quantile, where the premise is to select a subset of elements whose quantile approximates the true quantile. From this perspective, the latter body of research requires a certain amount of memory that increases as the required accuracy of the estimator increases \cite{Weide1978}.
Furthermore, in the case where the underlying distribution changes over time, those methods suffer from large bias in the summary information since the stored data might be stale \cite{Chen2000}. Examples of these works include \cite{Arasu2004, Weide1978, Munro1980, Greenwald2001, guha2009stream}. Guha and McGregor \cite{guha2009stream} advocate the use of random-order data models in contrast to adversarial-order models. They show that computing the median requires exponential number of passes in adversarial model while requiring $O (\log \log n)$ in random order model.

%\paragraph{Weight and dynamic environment}
In \cite{Chen2000, cao2010tracking, cao2009incremental, Chambers2006}, the authors proposed modifications of the stochastic approximation algorithm \cite{Tierney1983}. While Tierney \cite{Tierney1983} uses a sample mean update from previous quantile estimates, \cite{Chen2000, cao2010tracking, cao2009incremental, Chambers2006} propose an exponential decay in the usage of old estiamtes. This modification is particularly helpful to track quantiles of non-stationary data stream distributions. Indeed, a ``weighted'' update scheme is applied to incrementally build local approximations of the distribution function in the neighborhood of the quantiles. More recent approaches in this direction is the Frugal algorithm by Ma et al. \cite{ma2013frugal}, which is an additive alternative to the multiplicative estimator in Eq. \eqref{eq:1}, and the DQTRE and DQTRSE algorithms by Tiwari and Pandey \cite{tiwari2018technique}. A nice property of the DUMIQE in Eq. \eqref{eq:1} and the estimator suggested in this paper is that the update size is automatically adjusted dependent on the scale/range of the data. This makes the estimators robust to substantial changes in the data stream. The DQTRE and DQTRSE aims to achieve the same by estimating the range of the data using peak and valley detectors. However, a disadvantage with these algorithms is that several tuning parameters are required to estimate the range making the algorithms challenging to tune.
%{\color{black} It is worth emphasizing that the work in \cite{Chen2000} only uses the sign of the difference between the quantile estimate and current estimate in a similar manner to \cite{Tierney1983}. }

In many network monitoring applications, quantiles are key indicators
for monitoring the performance of the system. For instance, system administrators are interested in monitoring the $95\%$ quantile  of the response time of a web-server so that to hold it under a certain threshold. Quantile tracking is also useful for detecting abnormal events and in intrusion detection systems in general. However, the immense traffic volume of high speed networks impose some computational challenges: little storage and the fact that the computation needs to be ``one pass'' on the data.
It is worth mentioning that the seminal paper of Robbins and Monro \cite{robbins1951stochastic} which established the field of research called  ``stochastic approximation'' \cite{kushner2003stochastic} have included an incremental quantile estimator as a proof of concept of the vast applications of the theory of stochastic approximation.
An extension of the latter quantile estimator which first appeared as example in \cite{robbins1951stochastic} was further developed in \cite{joseph2004efficient} in order to handle the case of ``extreme quantiles''. Moreover, the estimator provided by Tierney \cite{Tierney1983} falls under the same umbrella of the example given in \cite{robbins1951stochastic}, and thus can be seen as an extension of it.

As Arandjelovic remarks \cite{arandjelovic2015two}, most quantile estimation algorithms are not single-pass algorithms and thus are not applicable for streaming data. On the other hand, the  single pass algorithms are concerned with the exact computation of the quantile and thus require a storage space of the order of the size of the data which is clearly an unfeasible condition in the context of big data stream.

Thus, we submit that all work on quantile estimation using more than one pass, or storage of the same order of the size of the observations seen so far is not relevant in the context of this paper.

When it comes to memory efficient methods that require a small storage footprint, histogram based methods form an important class. A representative work in this perspective is due to Schmeiser and Deutsch \cite{schmeiser1977quantile}. In fact, they proposed to use equidistant bins where the boundaries are adjusted online. Arandjelovic et al. \cite{arandjelovic2015two} use a different idea than equidistant bins by attempting to maintain bins in a manner that maximizes the entropy of the corresponding estimate of the historical data distribution. Thus, the bin boundaries are adjusted in an online manner.
Nevertheless, histogram based methods have problems addressing the problem of tracking quantiles of the current data stream distribution\cite{cao2009incremental} and are mainly used to recursively update quantiles for all data received so far. %In addition, they are prone to outliers that might corrupt the estimates of the distribution.

In \cite{naumov2007exponentially}, the authors propose a memory efficient method for simultaneous estimation of several quantiles using interpolation methods and a grid structure where each internal grid point is updated upon receiving an observation. The application of this approach is limited for stationary data. The approximation of the quantiles relies on using linear and parabolic interpolations, while the tails of the distribution are approximated using exponential curves. It is worth mentioning that the latter algorithm is based on the $P^2$ algorithm \cite{jain1985p}.

%A notable work treating simultaneous estimation of the quantiles using elements from the theory of stochastic approximation is due to Cao et al. \cite{Cao2009incrementalmultiple}. The authors resorted to interpolation by defining some type of distance between the interpolated quantiles so that to ensure no "crossing" between the monotonic quantile estimates. Nevertheless, the interpolation uses "the density" estimate as in \cite{Tierney1983} and in \cite{Chen2000}, which is an operation that increases the complexity. Furthermore, the latter methods \cite{Cao2009incrementalmultiple,Tierney1983,Chen2000} are prone to outliers that might corrupt the estimates of the distribution.

In \cite{jain1985p}, Jain et al. resort to five markers so that to track the quantile, where the markers correspond to different quantiles and the min and max of the observations. Their concept is similar to the notion of histograms, where each marker has  two measurements, its height and its position. By definition, each marker has some ideal position, where some adjustments are made to keep it in its ideal position by counting the number of samples exceeding the marker. In simple terms, for example, if the marker corresponds to the $80\%$ quantile, its ideal position will be around the point corresponding to $80\%$ of the data points below the marker. However, such approach does not handle the case of non-stationary quantile estimation as the position of the markers will be affected by stale data points. Then based on the position of the markers, quantiles are computed by supposing that the curve passing through three adjacent markers is parabolic and by using a piecewise parabolic prediction function.

It is worth mentioning that an important research direction that has received little attention in the literature revolves around updating the quantile estimates under the assumption that portions of the data are deleted. Such assumption is realistic in many real life settings where data needs to be deleted due to the occurrence of errors, or because the data samples are merely out-of-date and thus should be replaced. The deletion triggers a re-computation of the quantile \cite{Cao2009incrementalmultiple}, which is considered a complex operation. Note that the case of deleted data is more challenging than the case of insertion of new data.
In fact, the insertion can be handled easily using either sequential or batch updates, while quantile update upon deletion requires more complex forms of updates.

Finally, Lou et al. \cite{luo2016quantiles} perform extensive experiments to compare several of the algorithms described above.

\section{Quantile Estimator Using a Generalized Exponentially Weighted Average of Observations}
\label{sec:ewqe}

Let $X_n$ denote a stochastic variable representing the possible outcomes from a data stream at time $n$ and let $x_n$ denote a random sample (realization) of $X_n$. We assume that $X_n$ is distributed according to some distribution $f_n(x)$ that varies dynamically over time $n$. We denote the cumulative distribution of $X_n$ with $F_n(x)$, i.e. $P(X_n \leq x) = F_n(x)$. Further, let $Q_{n}(q)$ denote the quantile associated with probability $q$, i.e $P(X_n \leq Q_n(q)) = F_n(Q_n(q)) = q$.

A weakness of the state-of-the-art DUMIQE in Eq. \eqref{eq:1} is that the update step size is independent of the amount of the current error in the quantile estimate. We now propose an incremental quantile estimator where the update step size is \textit{proportional} to the distance between the received sample and current estimate. Thus, if the current estimate is off-track compared to the data stream, the estimator will initiate large jumps to rapidly get back on-track. The suggested estimator is described formally as follows
\begin{align}
  \label{eq:12a}
  \begin{split}
    \widehat{Q}_{n+1}(q) &\leftarrow \widehat{Q}_{n}(q) + \lambda c_n \frac{q}{\mu_n^+ - \widehat{Q}_{n}(q)} \left|x_n - \widehat{Q}_{n}(q) \right| \hspace{10mm} \text{ if } x_n > \widehat{Q_{n}}(q) \\[2mm]
    \widehat{Q}_{n+1}(q) &\leftarrow \widehat{Q}_{n}(q) - \lambda c_n \frac{1-q}{\widehat{Q}_{n}(q) - \mu_n^-} \left|x_n - \widehat{Q}_{n}(q) \right| \hspace{10mm} \text{ if } x_n \leq \widehat{Q_{n}}(q)
   \end{split}
\end{align}
where $\mu^+ = E(X_n|X_n > \widehat{Q}_{n}(q))$ and $\mu^- = E(X_n|X_n < \widehat{Q}_{n}(q))$. Naturally, the conditional expectations satisfy the inequality
\begin{align*}
\mu^- < \widehat{Q}_{n}(q) < \mu^+
\end{align*}
such that $\mu_n^+ - \widehat{Q}_{n}(q) > 0$ and $\widehat{Q}_{n}(q) - \mu_n^- > 0$. The factors $q/(\mu_n^+ - \widehat{Q}_{n}(q))$ and $(1-q)/(\widehat{Q}_{n}(q) - \mu_n^-)$ are included to ensure that the estimator converges to the true quantile value.

The constants $c_n$ can be any sequence of positive and bounded values. The estimator performed well when the fractions in Eq. \eqref{eq:12a} were ``normalizied'' as follows
\begin{align}
  \label{eq:30}
  c_n = \left( \frac{q}{\mu_n^+ - \widehat{Q}_{n}(q)} + \frac{1-q}{\widehat{Q}_{n}(q) - \mu_n^-} \right )^{-1}
\end{align}
Substituting Eq. \eqref{eq:30} into Eq. \eqref{eq:12a} we get
\begin{align}
  \label{eq:12b}
  \begin{split}
   \widehat{Q}_{n+1}(q) &\leftarrow \widehat{Q}_{n}(q) + \lambda a_n \left|x_n - \widehat{Q}_{n}(q) \right|  \hspace{16mm}\text{ if } x_n > \widehat{Q_{n}}(q) \\
   \widehat{Q}_{n+1}(q) &\leftarrow \widehat{Q}_{n}(q) - \lambda (1-a_n) \left|x_n - \widehat{Q}_{n}(q) \right|   \hspace{6mm}\text{ if } x_n \leq \widehat{Q_{n}}(q)
   \end{split}
\end{align}
where
\begin{align}
  \label{eq:3}
  a_n = \frac{q}{\mu_n^+ - \widehat{Q}_{n}(q)} \left/ \left( \frac{q}{\mu_n^+ - \widehat{Q}_{n}(q)} + \frac{1-q}{\widehat{Q}_{n}(q) - \mu_n^-} \right ) \right.
\end{align}
Please note that since $\mu_n^+ - \widehat{Q}_{n}(q) > 0$ and $\widehat{Q}_{n}(q) - \mu_n^- > 0$ we have that $0 < a_n < 1$. By factoring out $\widehat{Q}_{n}(q)$ and $x_n$ we get
\begin{align*}
   \widehat{Q}_{n+1}(q) &\leftarrow (1 - \lambda a_n) \widehat{Q}_{n}(q) + \lambda a_n x_n  \hspace{26mm}\text{ if } x_n > \widehat{Q_{n}}(q) \\
   \widehat{Q}_{n+1}(q) &\leftarrow (1 - \lambda (1-a_n)) \widehat{Q}_{n}(q) + \lambda (1-a_n) x_n   \hspace{6mm}\text{ if } x_n \leq \widehat{Q_{n}}(q)
\end{align*}
which can be written as
\begin{align}
  \label{eq:2}
  \widehat{Q}_{n+1}(q) &\leftarrow (1 - b_n) \widehat{Q}_{n}(q) + b_n x_n
\end{align}
where $b_n = \lambda\left(a_n + I\left(x_n \leq \widehat{Q}_{n}(q)\right)(1-2a_n)\right)$ and $I(A)$ the indicator function returning one (zero) if $A$ is true (false).

Now we will present a theorem that catalogs the properties of the estimator $\widehat{Q}_{n}(q)$ for a stationary data stream, i.e. $X_n = X \sim F(x), \,\, n=1,2,\ldots$.
\begin{theorem}
\label{thm:1}
Let $Q(q) = F^{-1}(q)$ be the true quantile to be estimated. Applying the updating rule in Eq. \eqref{eq:2}, we obtain:
\begin{align*}
\lim_{n \lambda \to \infty, \lambda \to 0}  \widehat{Q}_{n}(q) = Q(q)
\end{align*}
\end{theorem}
\noindent The proof of the theorem can be found in Appendix \ref{app:proof}. Although the quantile estimator $\widehat{Q}_{n}(q)$ given in Eq. \eqref{eq:2} is designed to estimate quantiles for dynamic environments, it is an important requirement that the estimator converges to the true quantile for static data streams as verified by Theorem \ref{thm:1}.

We end this section with a remark.\\
\textit{Remark 1:} If the conditional expectations are symmetrically positioned on each side of the quantile estimate, then $\widehat{Q}_{n}(q) - \mu_n^- = \widehat{Q}_{n}(q) - \mu_n^-$ and $a_n = q$ which is equal to DUMIQE. In other words, we can interpret that $\widehat{Q}_{n}(q) - \mu_n^-$ and $\widehat{Q}_{n}(q) - \mu_n^-$ ensure that the update rules take into account the asymmetries of the data stream distribution on each side of the quantile.

\subsection{Connection to the EWA}
\label{sec:connecEWA}

A simple and intuitive approach to track the expectation of a data stream distribution, i.e. $\mu_n = E(X_n)$, is the weighted moving average
\begin{align}
\label{eq:19}  \widehat{\mu}_n = \frac{1}{W_n} \sum_{i=0}^n w_i x_i
\end{align}
where $W_n = \sum_{j=1}^nw_j$. Using $w_{n-h} = \cdots = w_n = 1$ and the other weights equal to zero, Eq. \eqref{eq:19} reduces to the standard moving average. Intuitively, it seems more reasonable to use weights with decreasing values. The decrease should be more rapid than the standard sample mean $w_i = 1/i$ to be able to track the changes in the data stream.

Consider the following recursive update scheme
\begin{align}
\label{eq:20}  \widehat{\mu}_0 &\leftarrow x_0 \\
\label{eq:21}  \widehat{\mu}_{n+1} &\leftarrow (1 - \alpha) \widehat{\mu}_{n} + \alpha x_n
%\label{eq:22}                  &= \widehat{\mu}_{n} + \alpha(x_n - \widehat{\mu}_{n})
\end{align}
where the current estimate is a convex combination of the estimate at the previous time step and the observation. By substitution, we get
\begin{align}
  \label{eq:40}
  \widehat{\mu}_{n+1} &= \alpha (x_n + (1-\alpha)x_{n-1} + (1-\alpha)^2x_{n-2} + \cdots + (1-\alpha)^{n-1}x_{1}) + (1-\alpha)^n x_0
\end{align}
Interestingly, from Eq. \eqref{eq:40} we see that Eq. \eqref{eq:20} to Eq. \eqref{eq:21} can be interpreted as an EWA of observations. The estimator is highly popular and known to be the state-of-the-art approach to track expectations of dynamically varying data streams. Inspecting the incremental update form of our quantile estimator in Eq. \eqref{eq:2}, we see that it is identical to the update form of Eq. \eqref{eq:21}, except that the $0 < b_n < 1$ varies with time. Thus by keeping the weights constant as in Eq. \eqref{eq:21}, the estimator will track the expectation of the data stream distribution, while using the weights $0 <b_n < 1$ in Eq. \eqref{eq:2}, the estimator will track a quantile of the distribution.

\section{Quantile Estimation Algorithm}
\label{sec:qea}

The interpretation of the update rule in Eq. \eqref{eq:2} as an EWA of observations (recall Section \ref{sec:connecEWA}) and Theorem \ref{thm:1} constitute some intriguing theoretical results on the link between EMA and quantile estimation. However, the update rule in Eq. \eqref{eq:2} cannot be used directly since the conditional expectations, $\mu_n^+$ and $\mu_n^-$, are unknown and need to be estimated. Probably the most natural approach is to track conditional expectations using an EWA of observations as given in in Eq. \eqref{eq:20} to Eq. \eqref{eq:21}. This results in the following update rules: 
\begin{align}
  &\bullet\hspace{4mm} \label{eq:8}\widehat{Q}_{n+1}(q) \leftarrow (1 - \widehat{b}_n) \widehat{Q}_{n}(q) + \widehat{b}_n x_n \\[2mm]
  &\notag \bullet\hspace{4mm} \text{If } x_n > \widehat{Q_{n}}(q)\\
  &\hspace{8mm}\text{-}\hspace{4mm} \label{eq:6}\widehat{\mu}_{n+1}^+ \leftarrow \widehat{Q}_{n+1}(q) - \widehat{Q}_{n}(q) + (1-\gamma) \widehat{\mu}_{n}^+ + \gamma x_n\\
  &\hspace{8mm}\text{-}\hspace{4mm} \label{eq:6b}\widehat{\mu}_{n+1}^- \leftarrow \widehat{Q}_{n+1}(q) - \widehat{Q}_{n}(q) + \widehat{\mu}_{n}^- \\
  &\notag \bullet\hspace{4mm}\text{Else}\\
  &\hspace{8mm}\text{-}\hspace{4mm} \label{eq:7b}\widehat{\mu}_{n+1}^+ \leftarrow \widehat{Q}_{n+1}(q) - \widehat{Q}_{n}(q) + \widehat{\mu}_{n}^+ \\
  &\hspace{8mm}\text{-}\hspace{4mm} \label{eq:7}\widehat{\mu}_{n+1}^- \leftarrow \widehat{Q}_{n+1}(q) - \widehat{Q}_{n}(q) + (1-\gamma) \widehat{\mu}_{n}^- + \gamma x_n \\
  &\bullet\hspace{4mm} \widehat{a}_{n+1} \leftarrow \frac{q}{\widehat{\mu}_{n+1}^+ - \widehat{Q}_{n+1}(q)} \left/ \left( \frac{q}{\widehat{\mu}_{n+1}^+ - \widehat{Q}_{n+1}(q)} + \frac{1-q}{\widehat{Q}_{n+1}(q) - \widehat{\mu}_{n+1}^-} \right ) \right.\\
  &\bullet\hspace{4mm} \widehat{b}_{n+1} \leftarrow \lambda\left(\widehat{a}_{n+1} + I\left(x_n \leq \widehat{Q}_{n+1}(q)\right)(1-2\widehat{a}_{n+1})\right)
\end{align}
In each of the equations \eqref{eq:6} to \eqref{eq:7}, the part $\widehat{Q}_{n+1}(q) - \widehat{Q}_{n}(q)$ is included to ensure that the conditional expectation estimates are relative to the current quantile estimate $\widehat{Q}_{n+1}(q)$.
%The resulting interpretation of the EWA tuning parameter, $\gamma$, in Eq. \eqref{eq:6} and Eq. \eqref{eq:7} is to make sure that the conditional expectations $\widehat{\mu}_{n+1}^+$ and $\widehat{\mu}_{n+1}^-$ are estimated correctly relative to the current quantile estimate.
Thus Eq. \eqref{eq:8} tracks the overall trends of the dynamical data stream while Eq. \eqref{eq:6} to Eq. \eqref{eq:7} are responsible for estimating the conditional expectations \textit{relative} to the quantile estimate. Thus, for most dynamic data streams it is reasonable to use a value of the EWA tuning parameter, $\gamma$, that is on a smaller scale than $\lambda$ \cite{konda2004convergence}. This is verified in our experiments. In the rest of the paper, we denote this EWA quantile estimator approach for QEWA. We end this section with a remark.

\textit{Remark1:} We evaluated a second approach based on estimating the streaming distribution, $f_n(x)$, and computing the unknown conditional expectations from the estimated distribution. The streaming distribution where estimated by tracking several quantiles $Q_n(q_1), Q_n(q_2), \ldots, \ldots, Q_n(q_{K})$ and a linear spline were interpolated between the quantile estimates. However experiments showed that the QEWA approach performed better than this spline approach. The spline approach therefore is not followed any further in the paper.

\section{Experiments based on Synthetic Data}
\label{sec:se}

\begin{figure*}
  \centering
    \includegraphics[width = 0.9\textwidth]{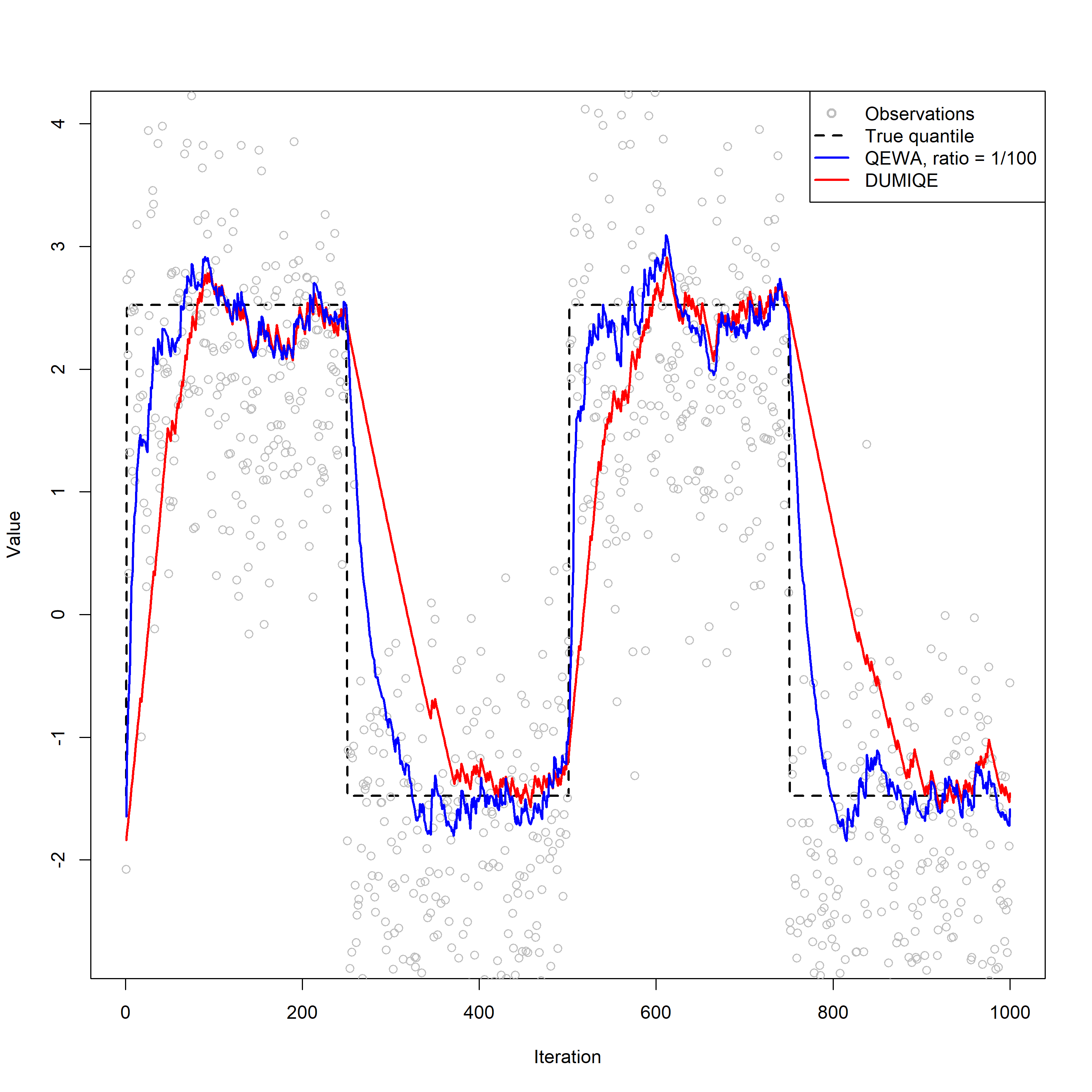}
  \caption{Quantile estimates in every iteration using the DUMIQE and the suggested algorithm QEWA using ratio $\gamma/\lambda = 1/100$.}
  \label{fig:10}
\end{figure*}
In this section we perform a thorough comparison of the performance of the suggested algorithm QEWA and other quantile estimators in the literature. Figure \ref{fig:10} shows tracking of the quantile with probability $q = 0.7$ for the suggested algorithm QEWA and DUMIQE. The true quantile is given as the dashed black line. The tuning parameters are adjusted such that the estimation error in the stationary parts after convergence is the same for the two algorithms. We see that the proposed algorithm QEWA tracks the true quantile more efficiently after a switch than the DUMIQE. For the suggested algorithm, the step size is proportional to the difference between the observations and the quantile estimate (recall Eq. \eqref{eq:12b}). After a switch, these differences are large, and our devised algorithm makes large steps to get back on-track. The DUMIQE, and the other state-of-the art incremental algorithms, use the same step size independent of the these difference, resulting in poorer tracking.

The results below show a more systematic evaluation of the performance of the suggested algorithm against seven state-of-the-art quantile estimators namely the DUMIQE and RUMIQE by Yazidi and Hammer \cite{yazidi17multiplicative}, the estimator due to Cao et al. \cite{cao2010tracking}, the Frugal approach by Ma et al. \cite{ma2013frugal}, the selection algorithm by Guha and McGregor \cite{guha2009stream} and the DQTRE and DQTRSE algorithms by Tiwari and Pandey \cite{tiwari2018technique}. For the DQTRE and DQTRSE algorithms we used values of the tuning parameters recommended in \cite{tiwari2018technique}, namely $\alpha = 0.1, \beta = (1 - \alpha)^{\lambda}, p_b = 1/10$ and $l = 1/4$ which performed well in our experiments.

The estimator in this paper is designed to perform well for dynamically changing data streams and the experiments will focus on such streams.
%It would have been interesting to evaluate the performance of the different methods for real life data, but this is challenging to do in a systematic way for dynamical data streams as the ground truth generally is missing. %Before proceeding to systematic experiments based synthetic data, we just recall Figure \ref{fig:1} showing that the quantile estimators presented in this paper can be used to efficiently track quantiles of challenging real life data streams.
%For static data streams, different online estimators can be compared to quantile estimates using state of the art off line estimators. Unfortunately, this approach does not generalize to dynamical systems and comparison between methods for real data is difficult in dynamical systems. Therefore, we will rather focus on performing thorough synthetic experiments.

We consider four different cases where we assume that the data are outcomes from a normal distribution or from a $\chi^2$ distribution. For two of the cases we look at both a case where the data stream varies smoothly (periodic) or switches rapidly (switch). For the normal distribution periodic case, we assume that the expectation of the distribution varies with time
\begin{align*}
\mu_n = a \sin \left( \frac{2\pi}{T} n \right), \,\,\, n = 1,2,3, \ldots
\end{align*}
which is the sinus function with period $T$. For the switch case, the expectation jumps between values $a$ and $-a$.
\begin{align*}
  %\label{eq:17}
  \mu_n = \left \{
  \begin{array}{ll}
    a & \text{ if } n\,\text{mod}\,T \leq T/2 \\
    -a & \text{ else }
  \end{array}
  \right.
\end{align*}
We assume that the standard deviation of the normal distribution does not vary with time and is equal to one.

For the $\chi^2$ distribution periodic case, we assume that the number of degrees of freedom varies with time as follows
\begin{align*}
  \nu_n = a \sin \left( \frac{2\pi}{T} n \right) + b, \,\,\, n = 1,2,3, \ldots
\end{align*}
where $b > a$ such that $\nu_n > 0$ for all $n$. For the switch case, the number of degrees of freedom jumps between values $a + b$ and $-a + b$
\begin{align*}
  %\label{eq:18}
  \mu_n = \left \{
  \begin{array}{ll}
    a+b & \text{ if } n\,\text{mod}\,T \leq T/2 \\
    -a+b & \text{ else }
  \end{array}
  \right.
\end{align*}
In the experiments we used $a = 2$ and $b=6$. 

We estimated quantiles of both the normally and $\chi^2$ distributed data streams above using two different periods, namely $T=100$ (rapid variation) and $T=500$ (slow variation), i.e. in total eight different data streams. For each data stream we estimated the $50$, $70$ and $90\%$ quantiles ending up with a total of $24$ different estimation tasks.

To measure estimation error, we used the root mean squares error (RMSE) for each quantile given as:
\begin{align}
  \label{eq:27}
  \text{RMSE}\, =  \sqrt{ \frac{1}{N}\sum_{n=1}^N \left(Q_n(q) - \widehat{Q}_n(q)\right)^2 }
\end{align}
where $N$ is the total number of samples in the data stream. In the experiments, we used $N = 10^6$ which efficiently removed any Monte Carlo errors in the experimental results. In order to obain a good overview of the performance of the algorithms, we measured the estimation error for a large set of different values of the tuning parameters of the algorithms.

\begin{figure}
  \centering
  \begin{tabular}{cc}
   \includegraphics[width = 0.5\textwidth]{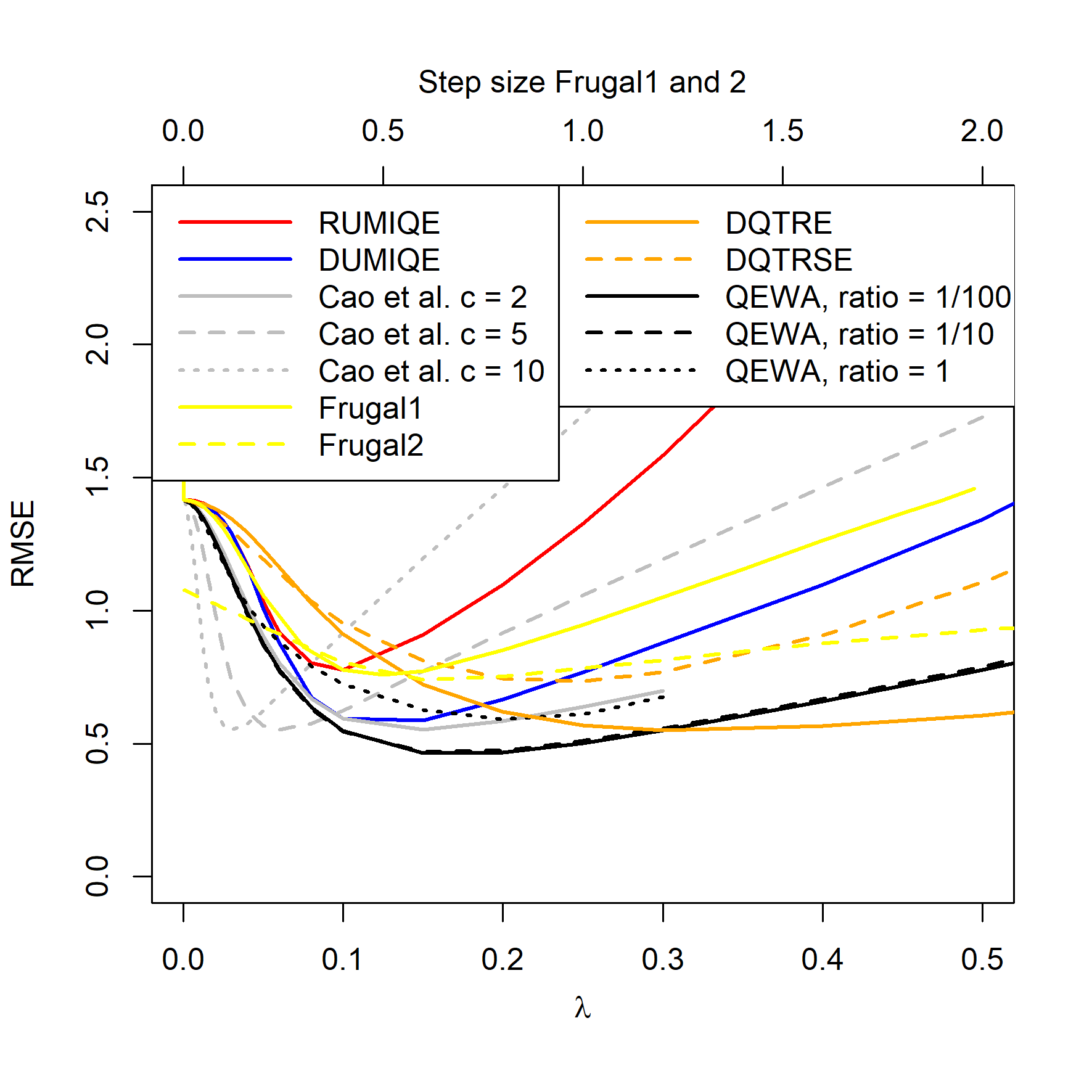} & \includegraphics[width = 0.5\textwidth]{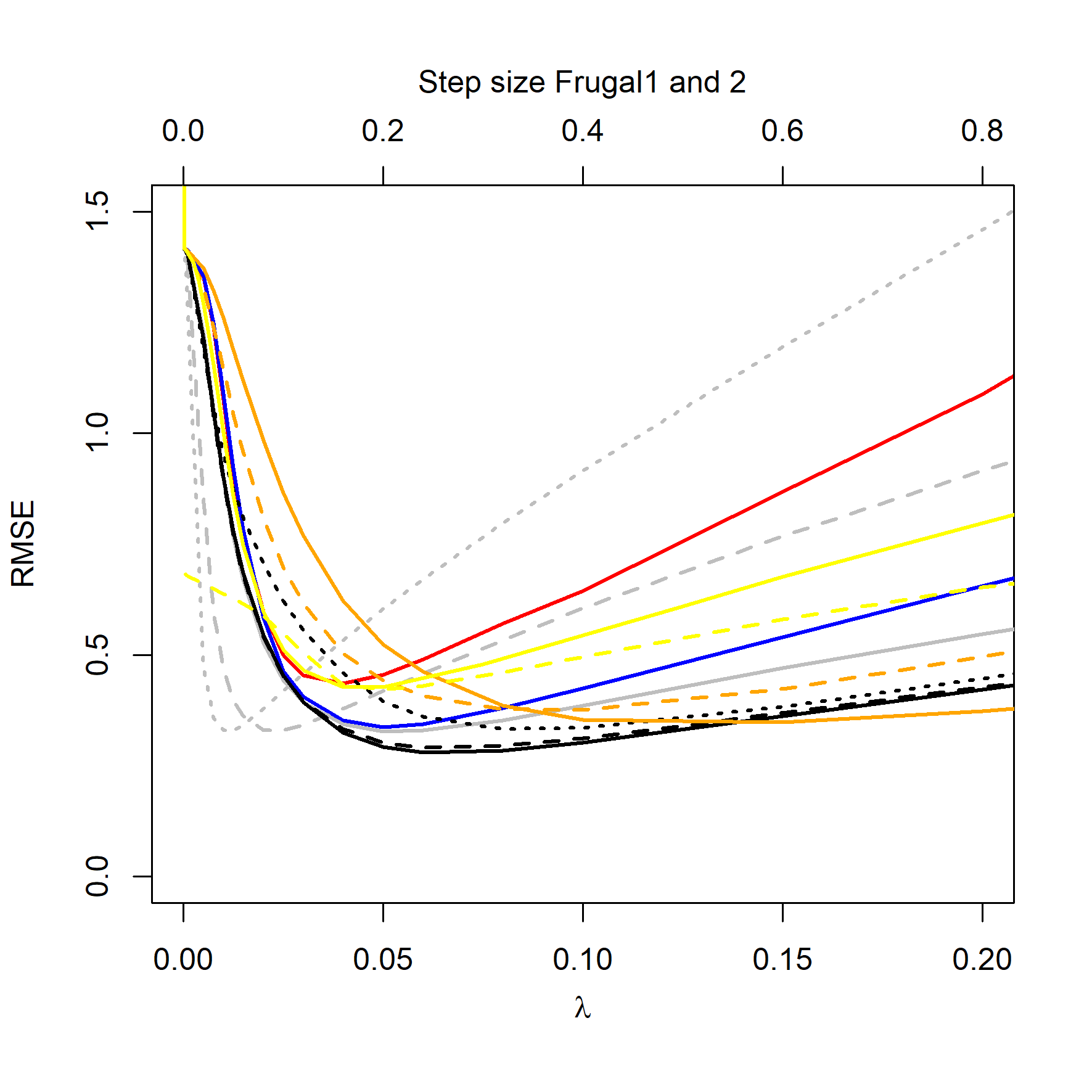} \\
   \includegraphics[width = 0.5\textwidth]{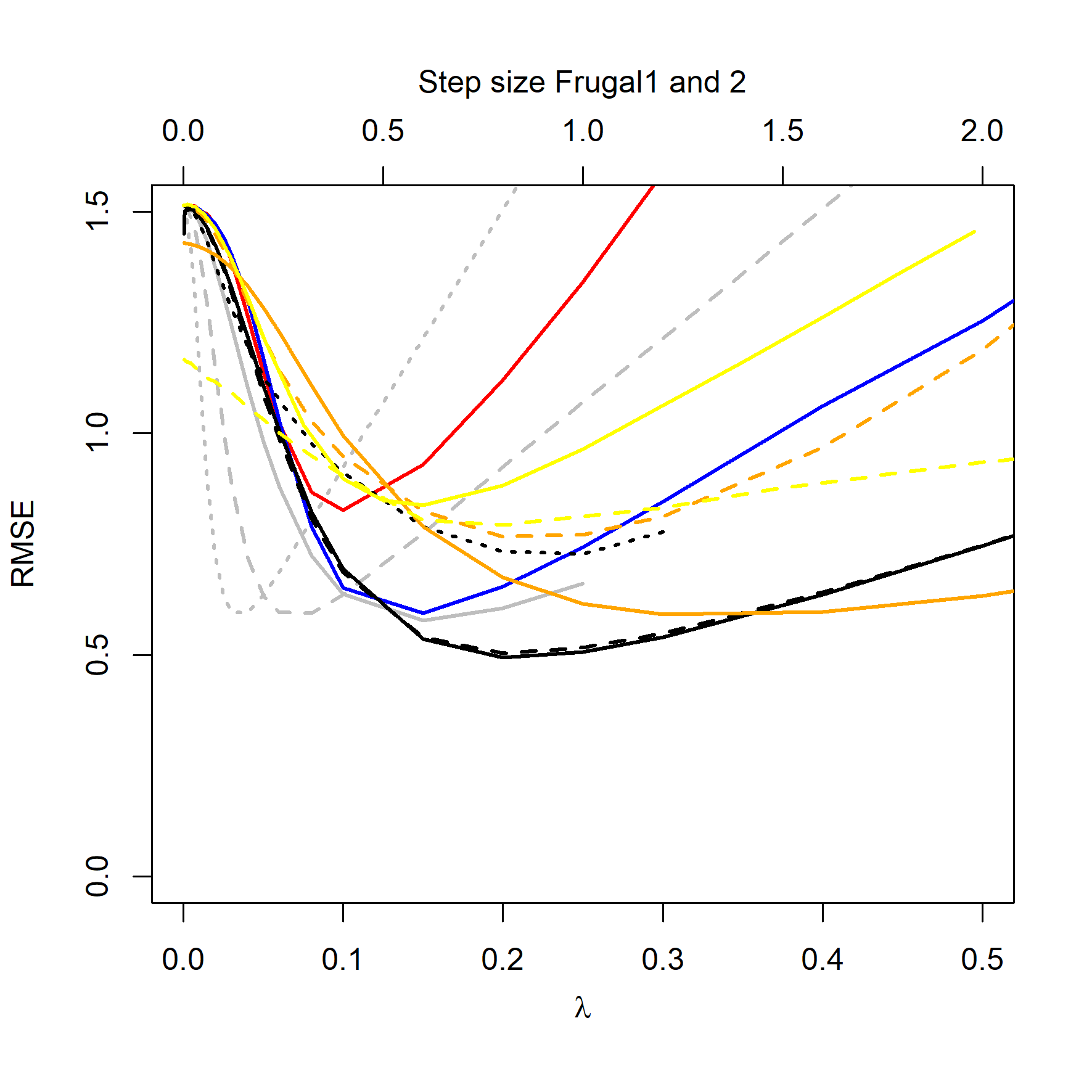} & \includegraphics[width = 0.5\textwidth]{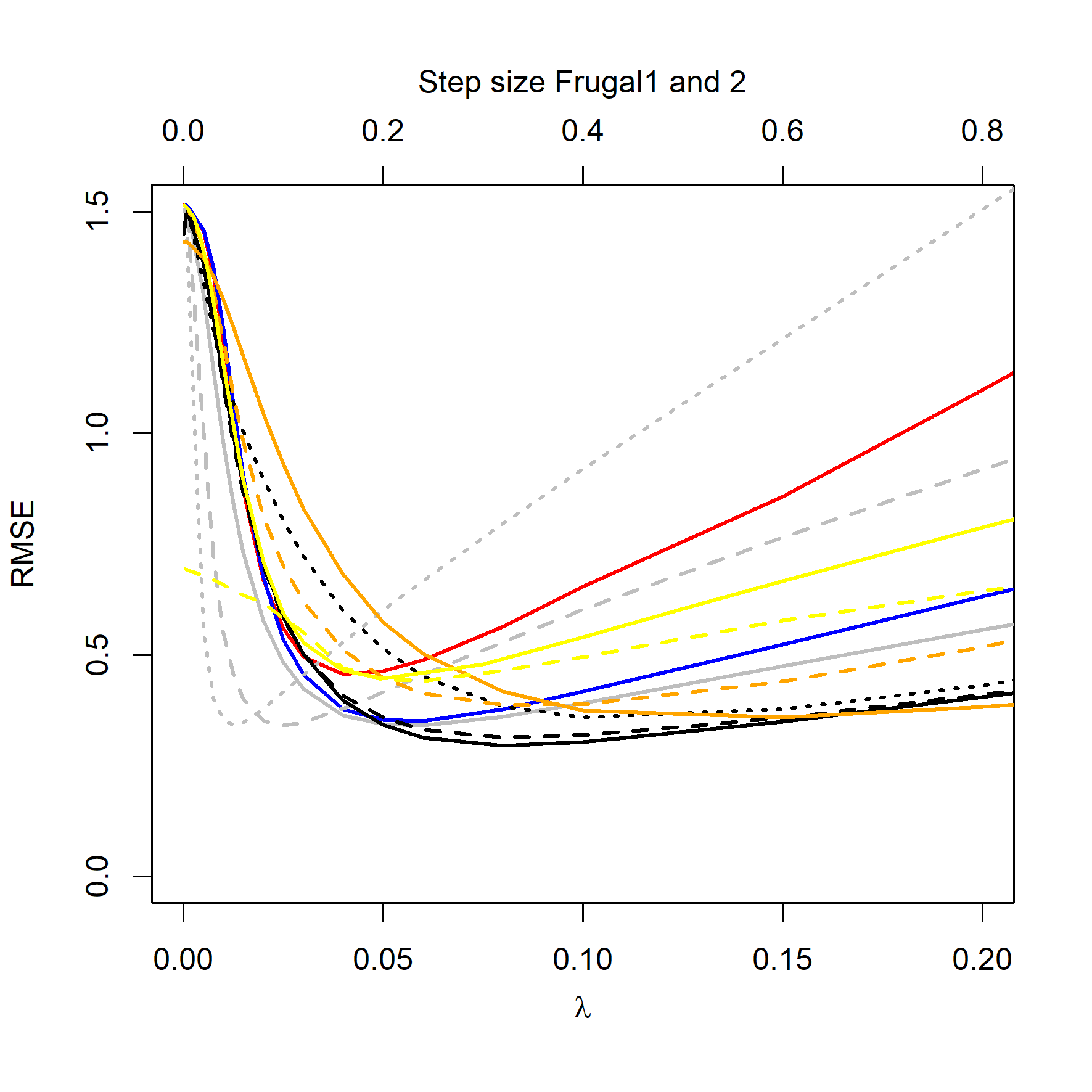} \\
   \includegraphics[width = 0.5\textwidth]{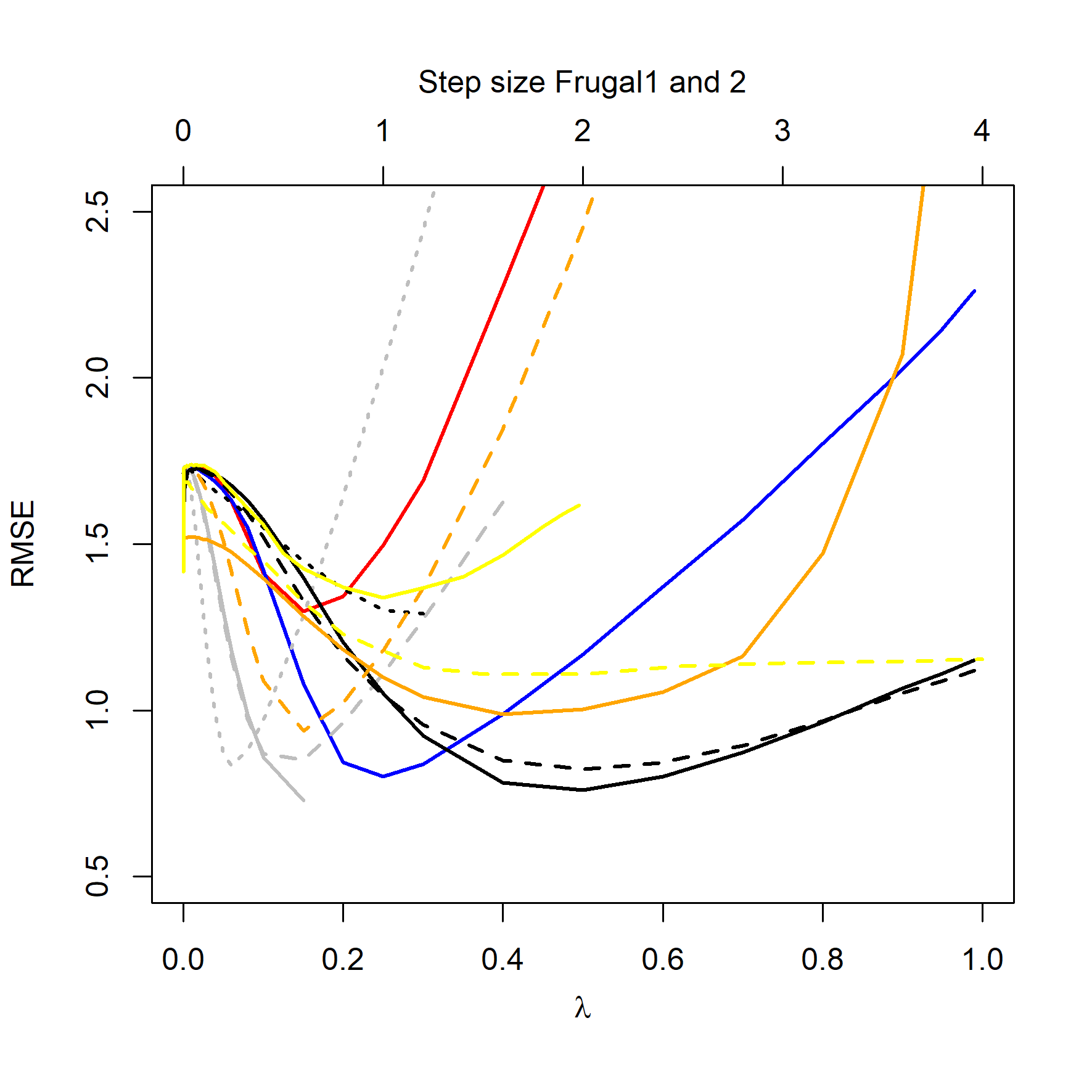} & \includegraphics[width = 0.5\textwidth]{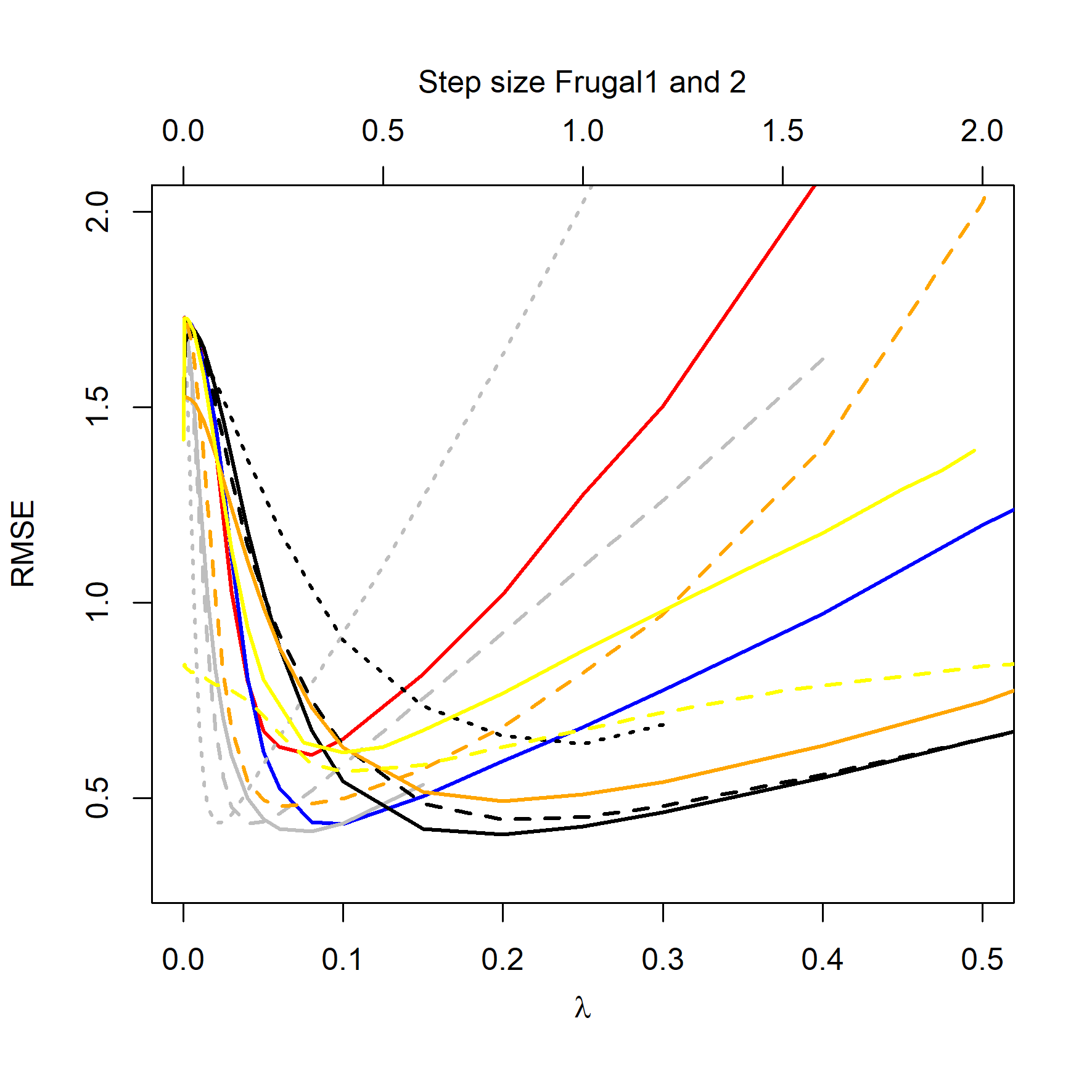}
  \end{tabular}
  \caption{Normal distribution periodic case: The left and right columns show results for $T=100$ and $T=500$, respectively. The rows from top to bottom show results when estimating quantile $Q_n(q=0.5),\,\, Q_n(q=0.7)$ and $Q_n(q=0.9)$, respectively. Ratio refers to the ratio between the tuning parameters, i.e. ratio = $\gamma/\lambda$. The upper $x$ axis refers to the step size in the Frugal algorithms.}
  \label{fig:2}
\end{figure}
\begin{figure}
  \centering
  \begin{tabular}{cc}
   \includegraphics[width = 0.5\textwidth]{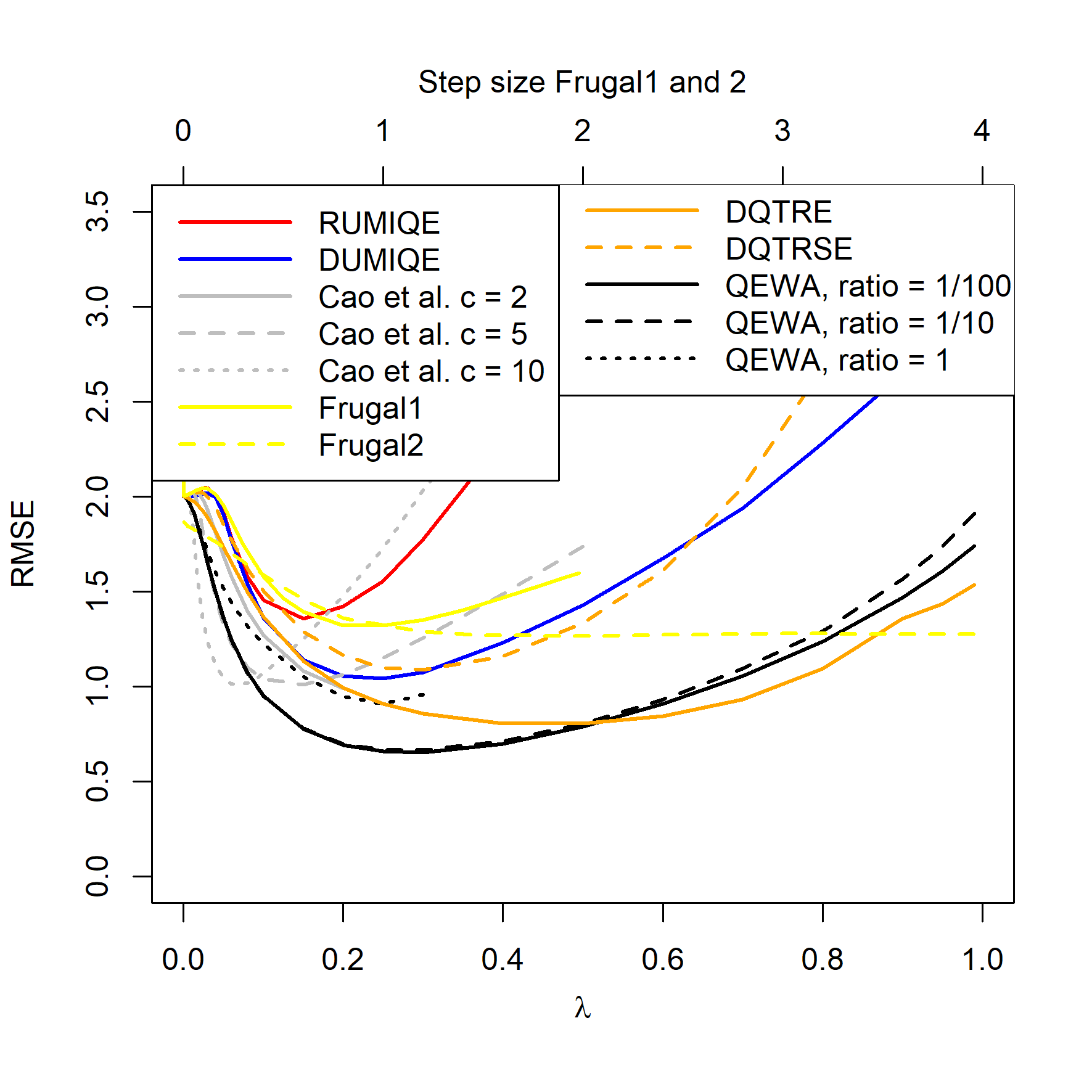} & \includegraphics[width = 0.5\textwidth]{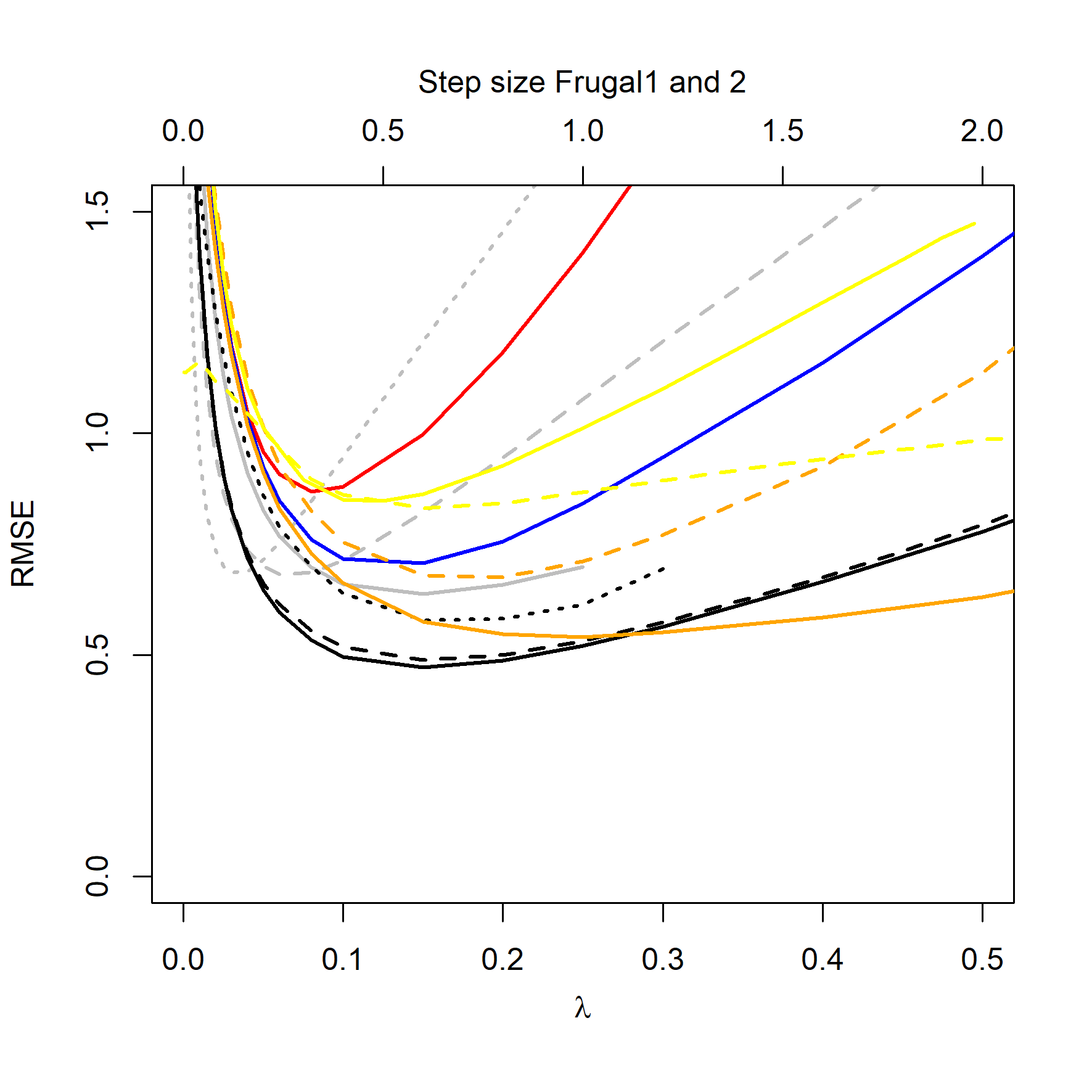} \\
   \includegraphics[width = 0.5\textwidth]{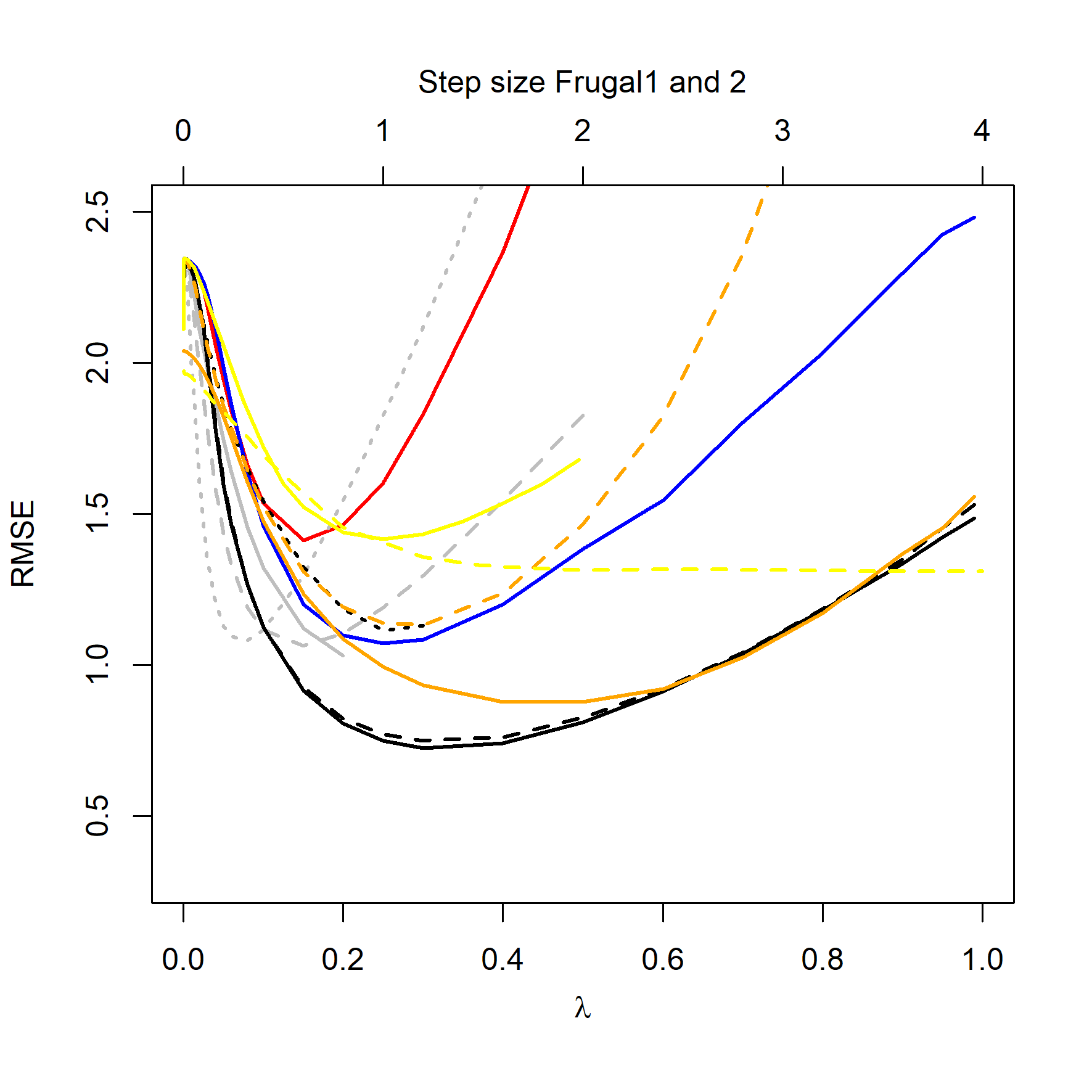} & \includegraphics[width = 0.5\textwidth]{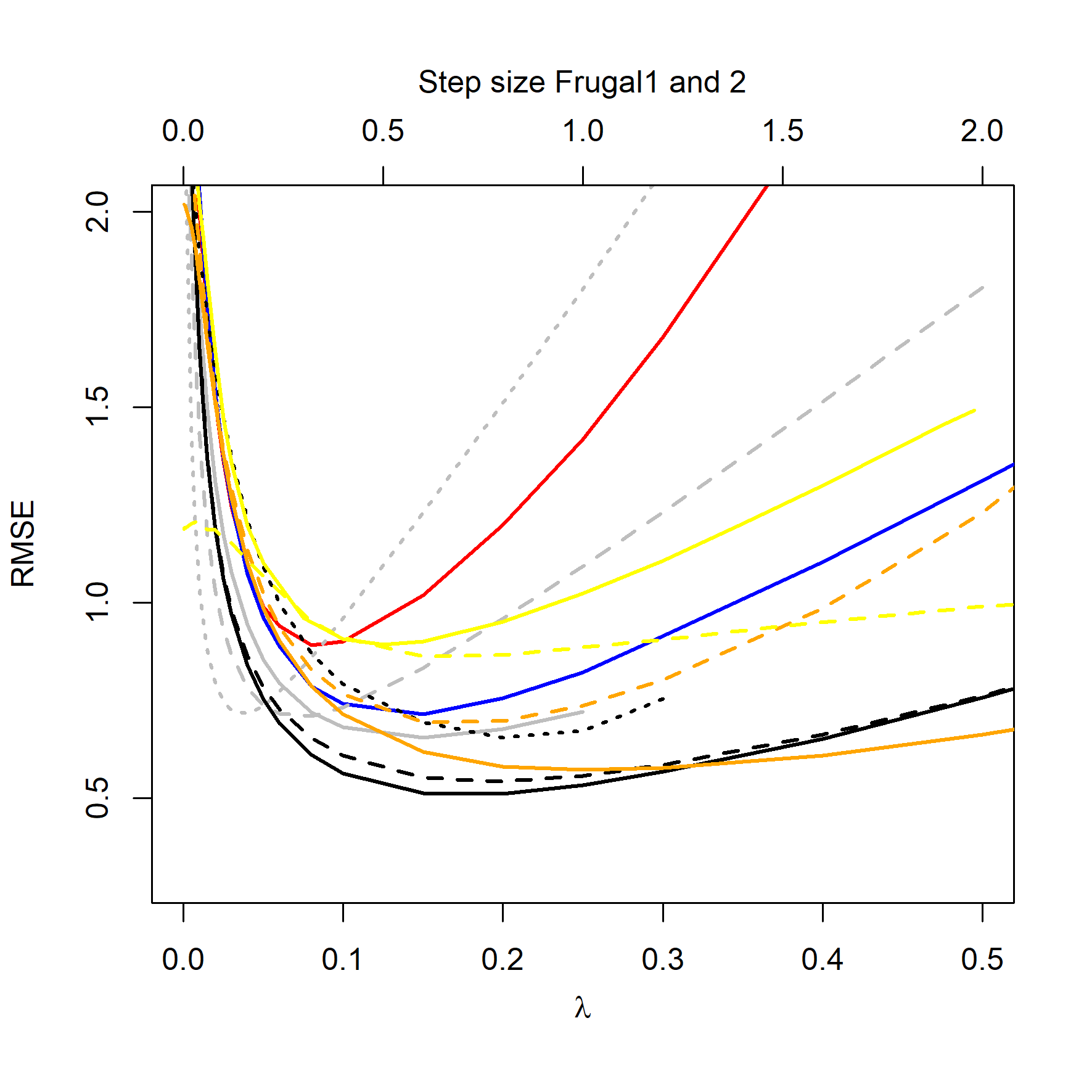} \\
   \includegraphics[width = 0.5\textwidth]{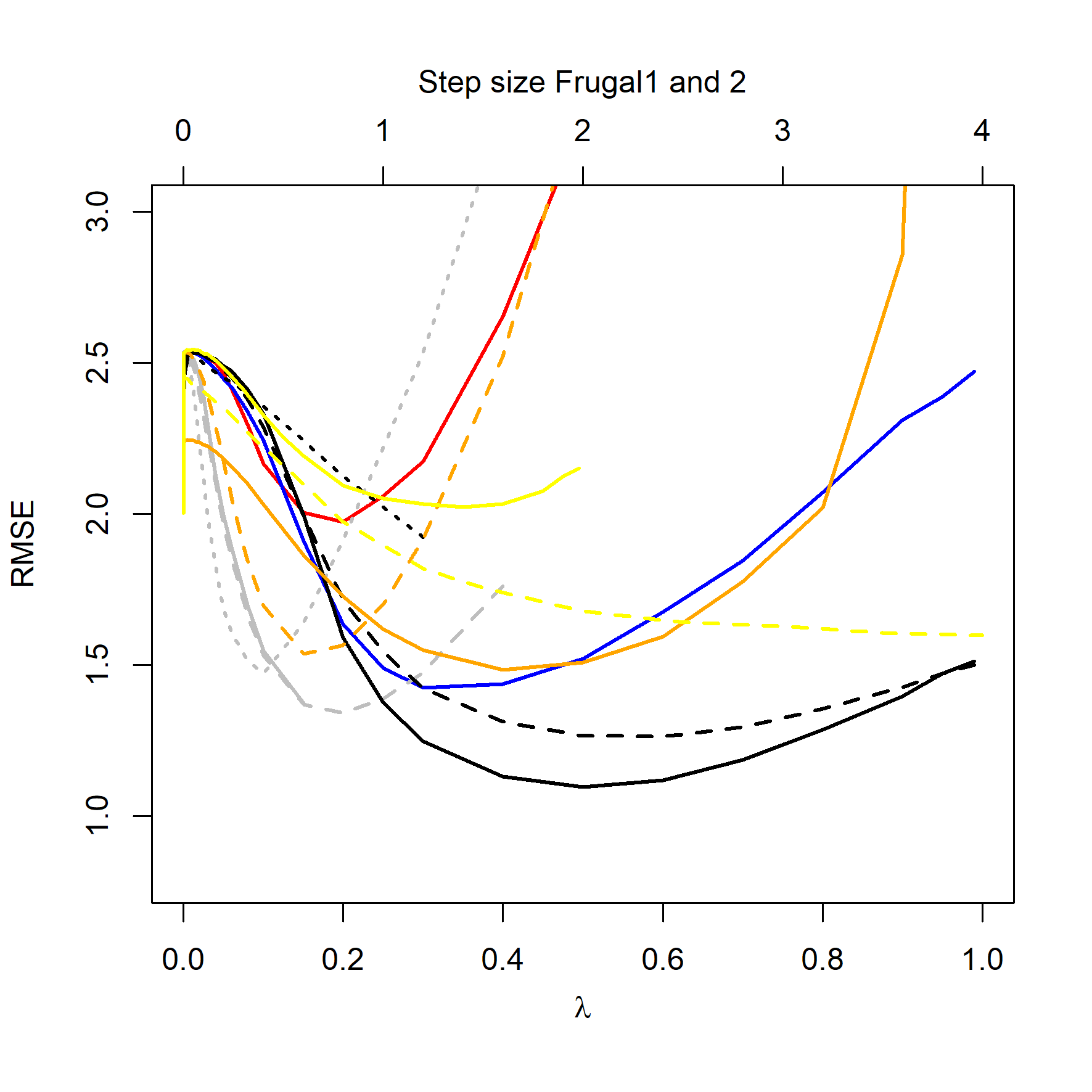} & \includegraphics[width = 0.5\textwidth]{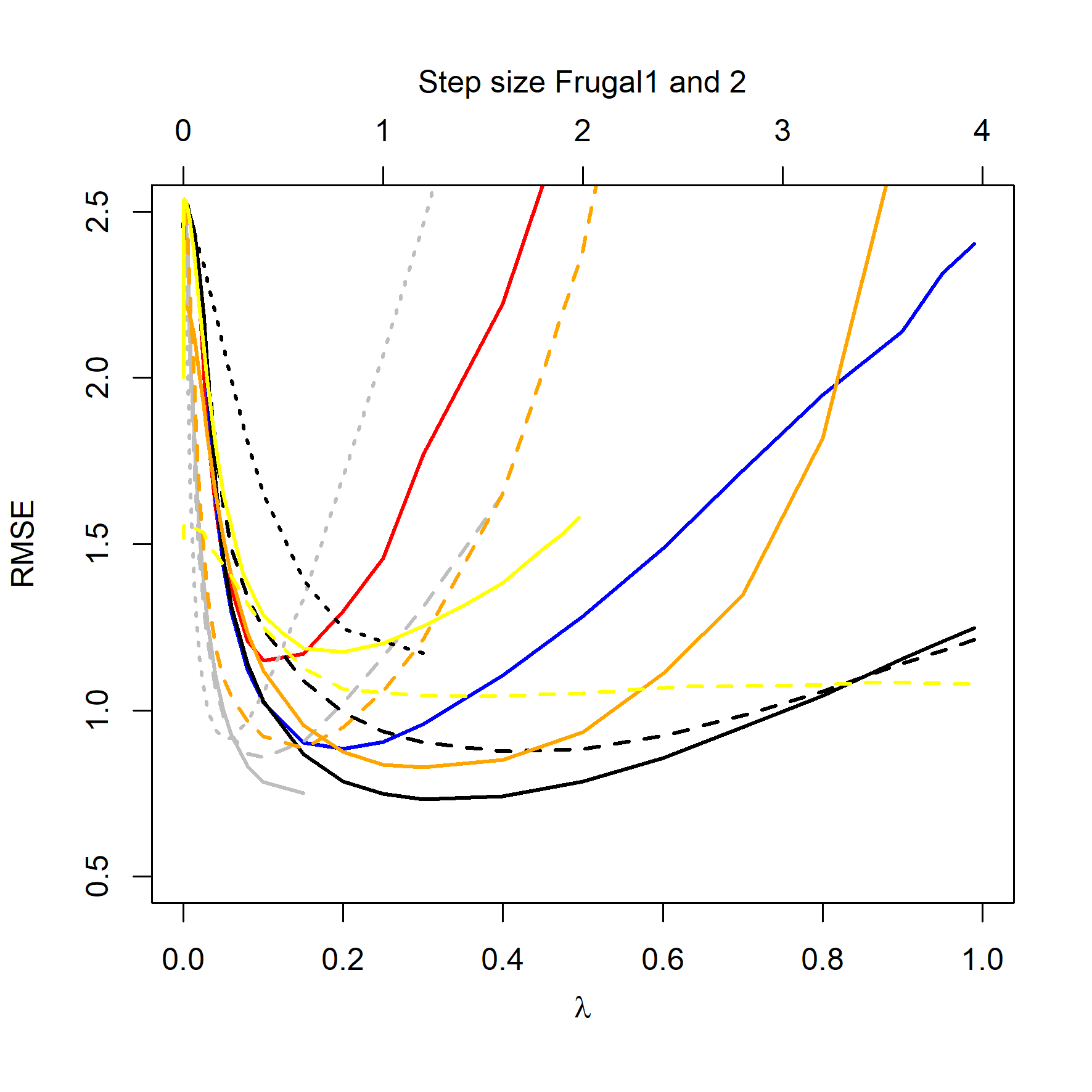}
  \end{tabular}
  \caption{Normal distribution switch case: The left and right columns show results for $T=100$ and $T=500$, respectively. The rows from top to bottom show results when estimating quantile $Q_n(q=0.5),\,\, Q_n(q=0.7)$ and $Q_n(q=0.9)$, respectively. Ratio refers to the ratio between the tuning parameters, i.e. ratio = $\gamma/\lambda$. The upper $x$ axis refers to the step size in the Frugal algorithms.}
  \label{fig:3}
\end{figure}
\begin{figure}
  \centering
  \begin{tabular}{cc}
   \includegraphics[width = 0.5\textwidth]{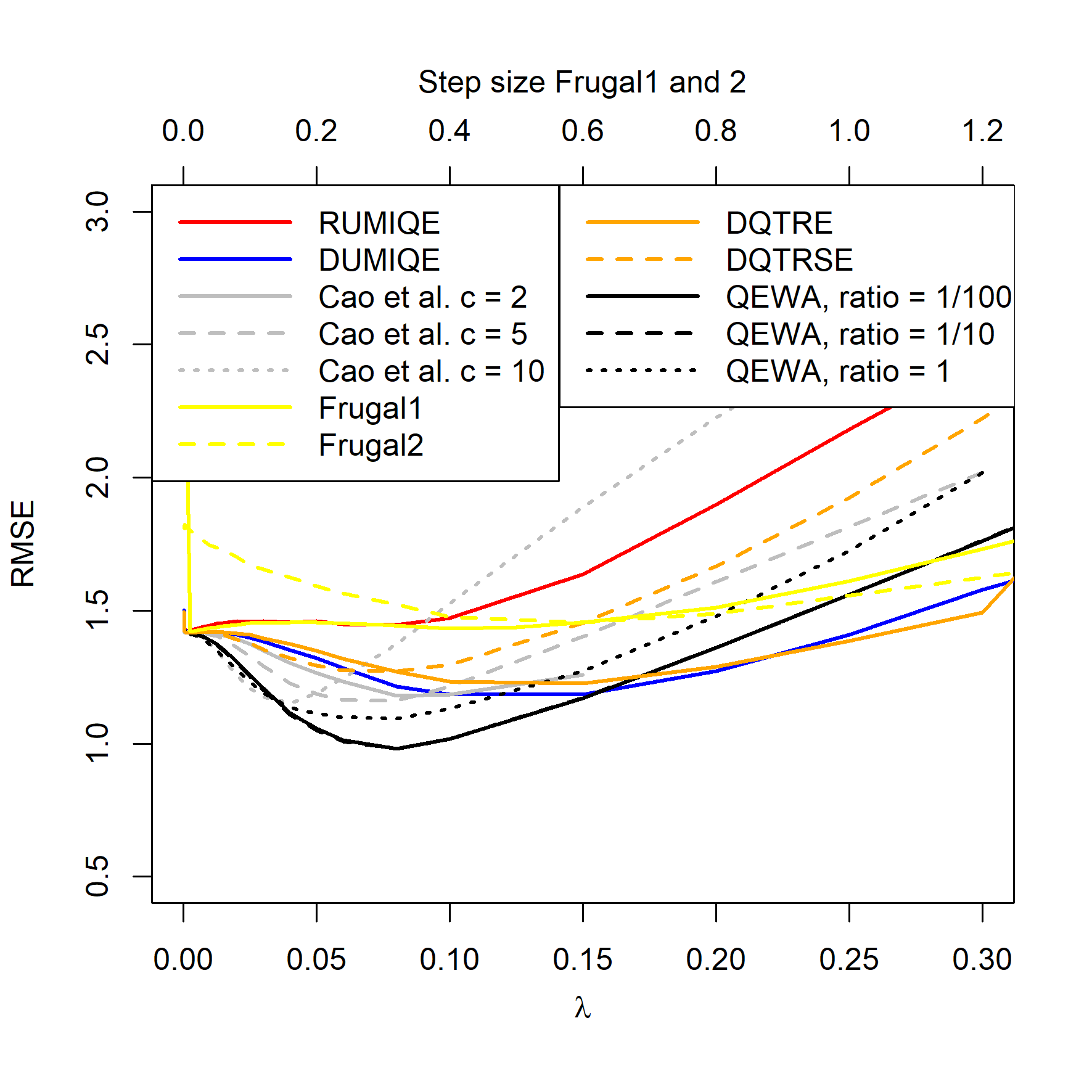} & \includegraphics[width = 0.5\textwidth]{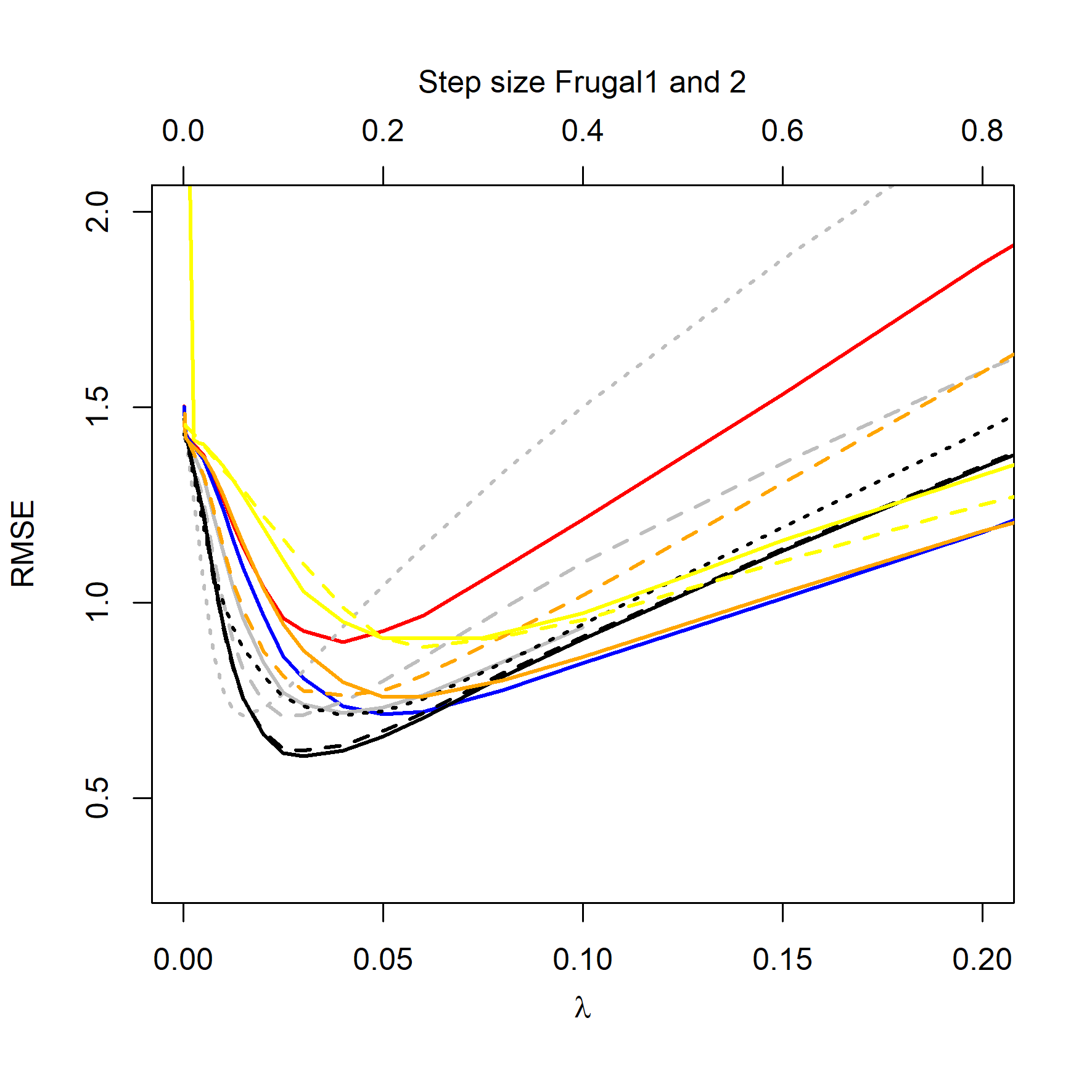} \\
   \includegraphics[width = 0.5\textwidth]{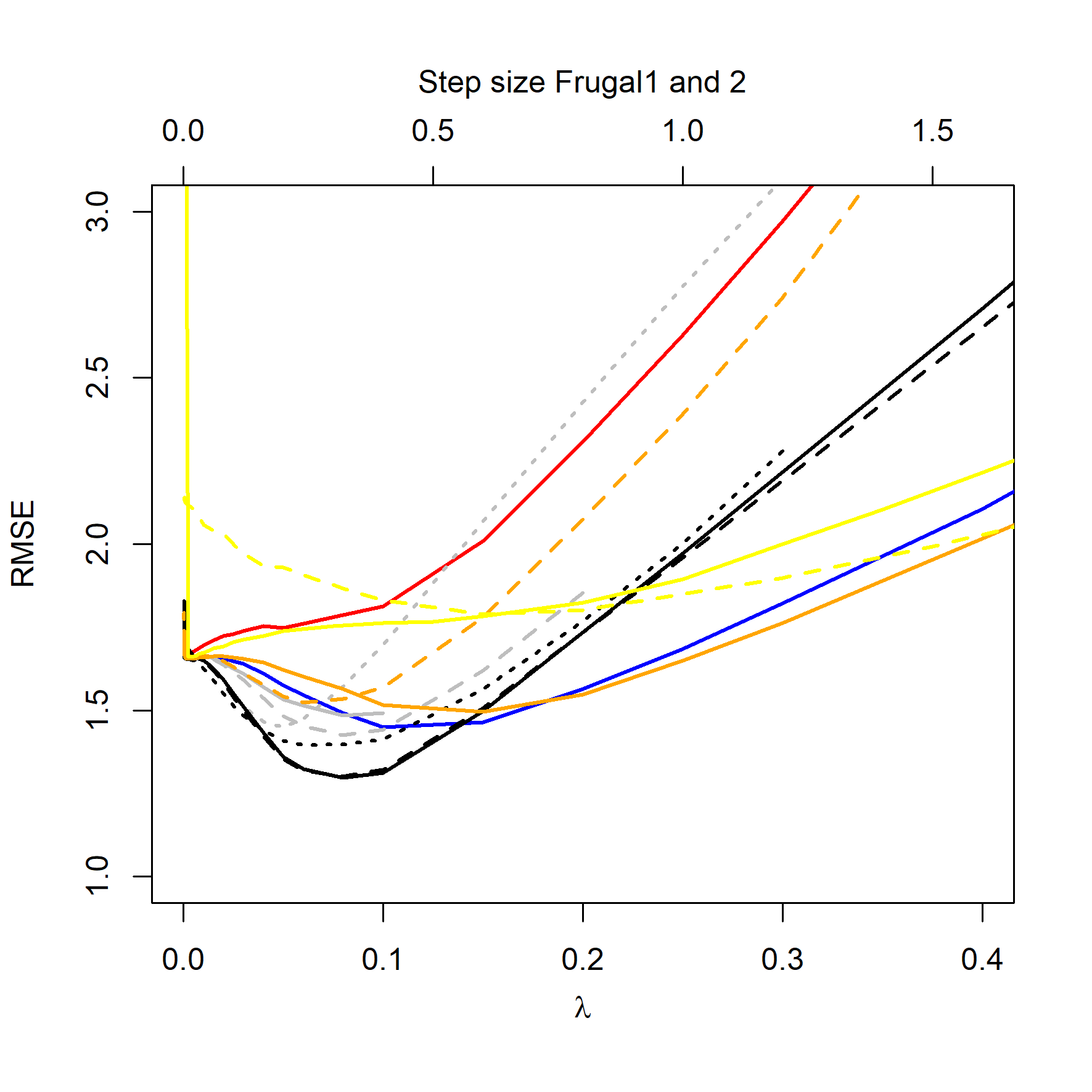} & \includegraphics[width = 0.5\textwidth]{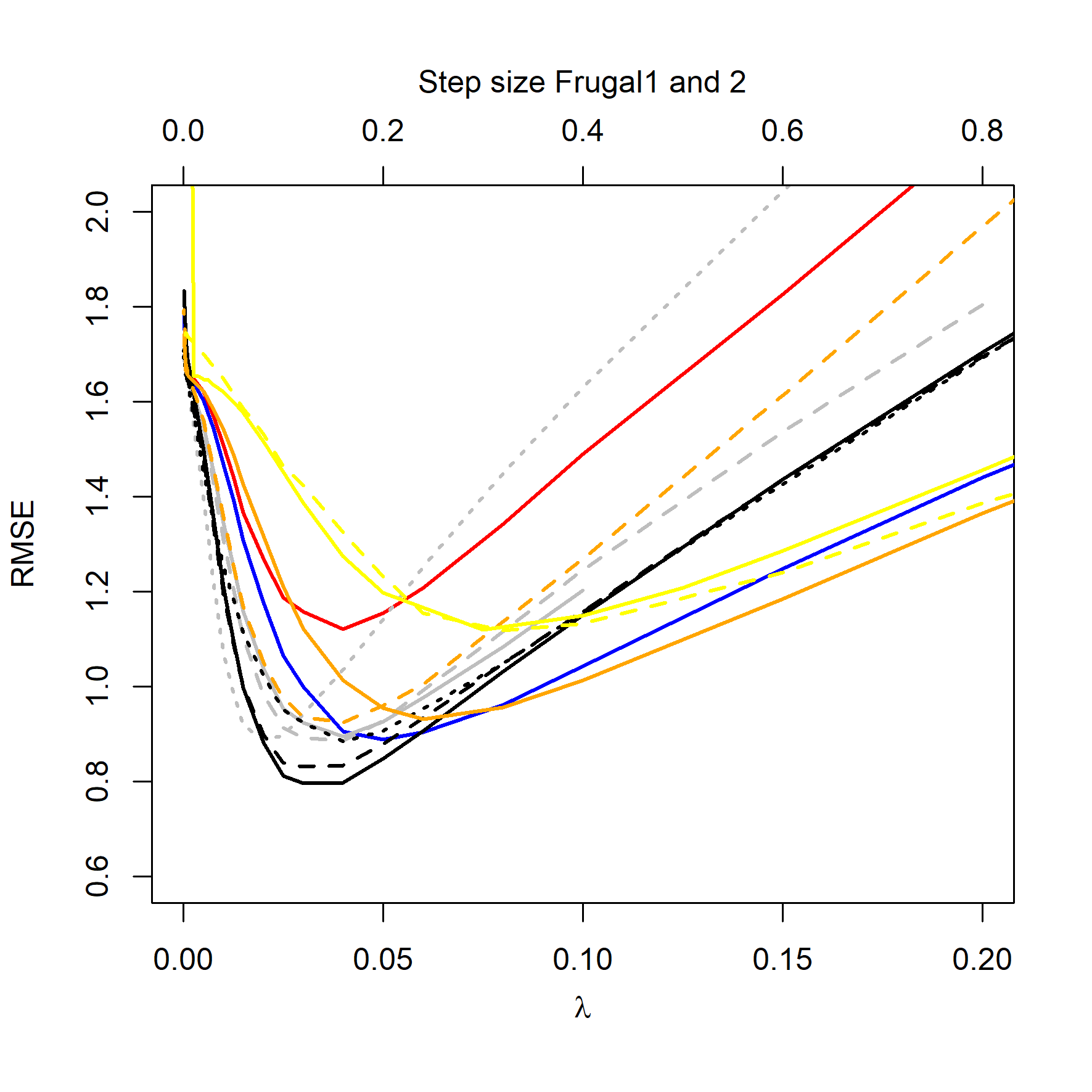} \\
   \includegraphics[width = 0.5\textwidth]{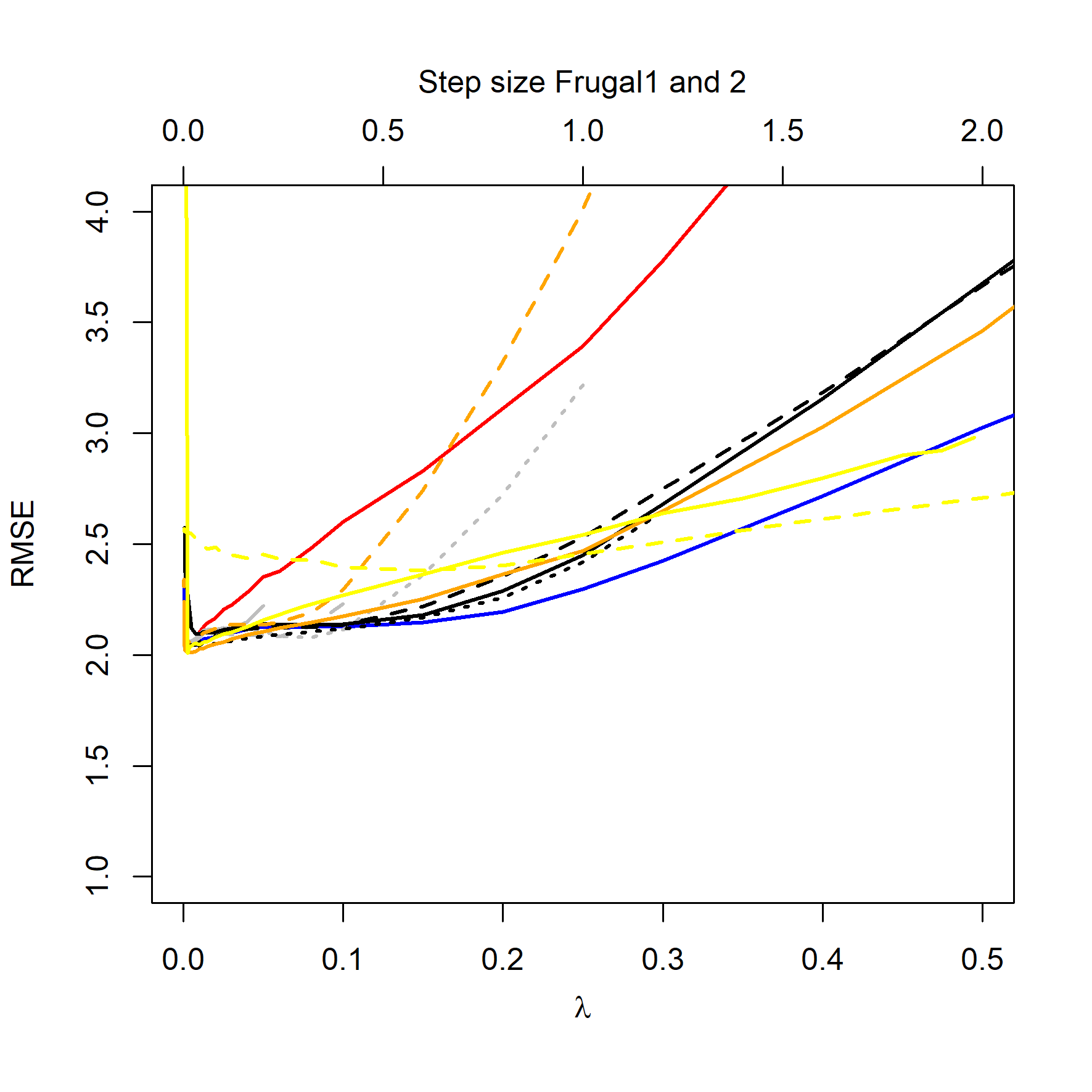} & \includegraphics[width = 0.5\textwidth]{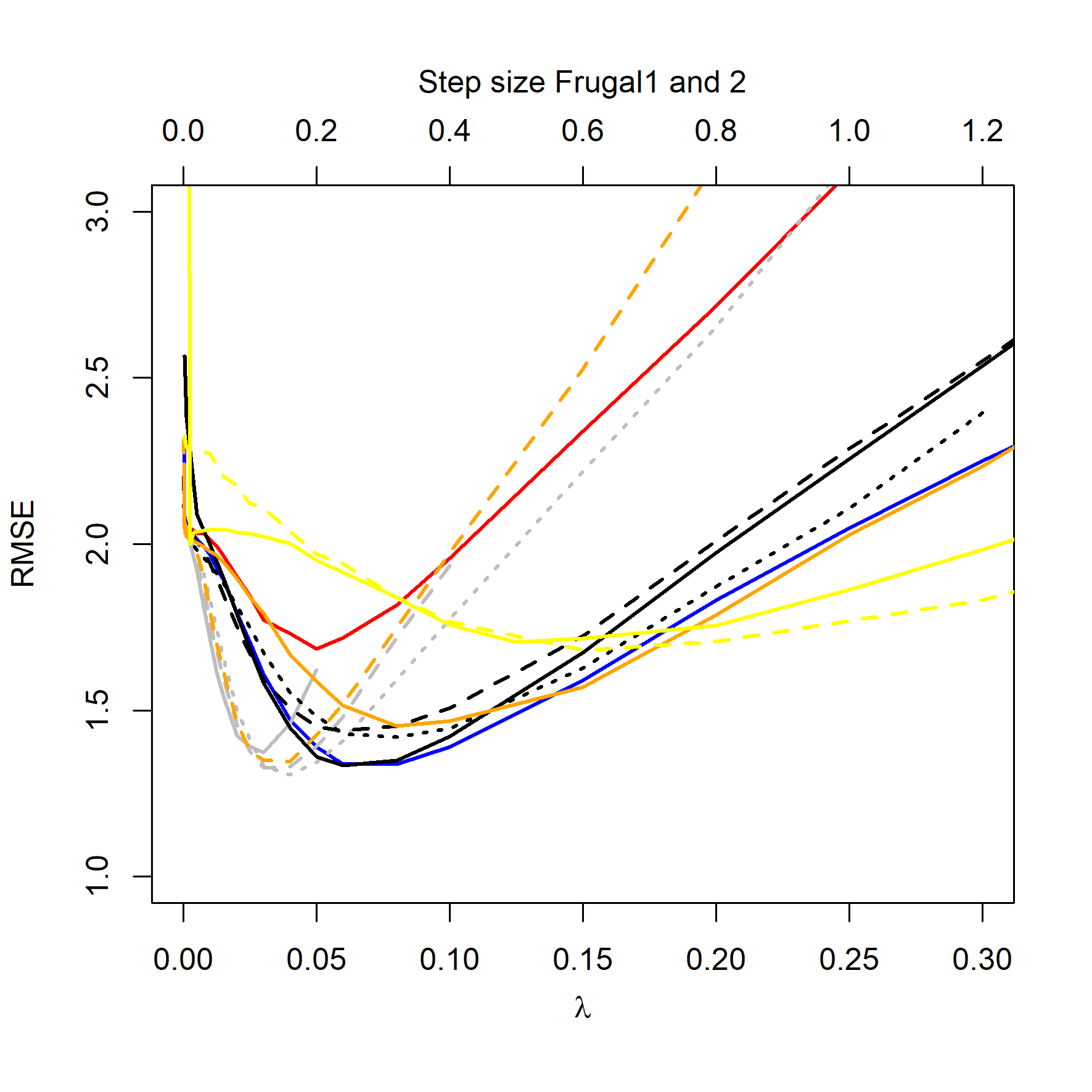}
  \end{tabular}
  \caption{$\chi^2$ distribution periodic case: The left and right columns show results for $T=100$ and $T=500$, respectively. The rows from top to bottom show results when estimating quantile $Q_n(q=0.5),\,\, Q_n(q=0.7)$ and $Q_n(q=0.9)$, respectively. Ratio refers to the ratio between the tuning parameters, i.e. ratio = $\gamma/\lambda$. The upper $x$ axis refers to the step size in the Frugal algorithms.}
  \label{fig:4}
\end{figure}
\begin{figure}
  \centering
  \begin{tabular}{cc}
   \includegraphics[width = 0.5\textwidth]{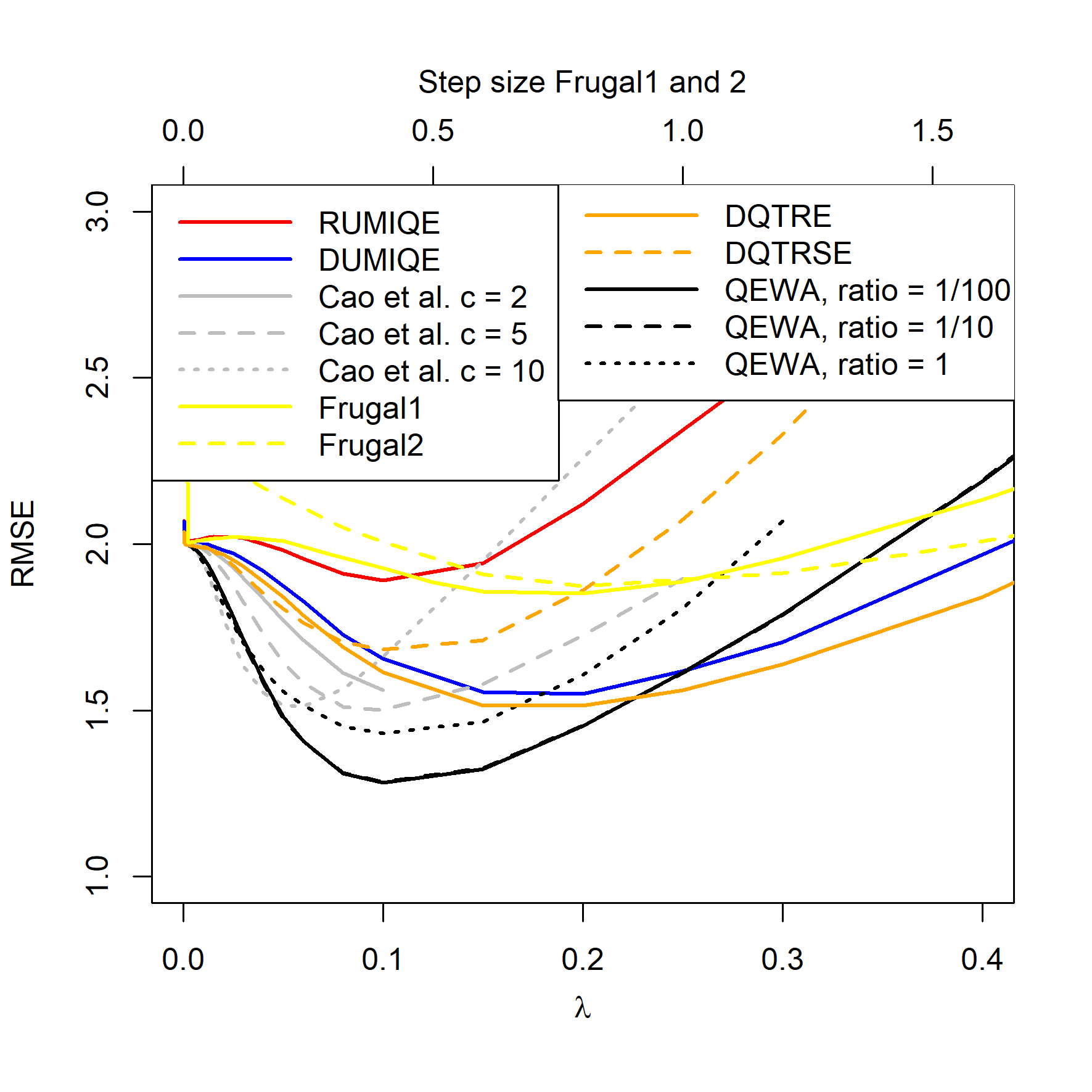} & \includegraphics[width = 0.5\textwidth]{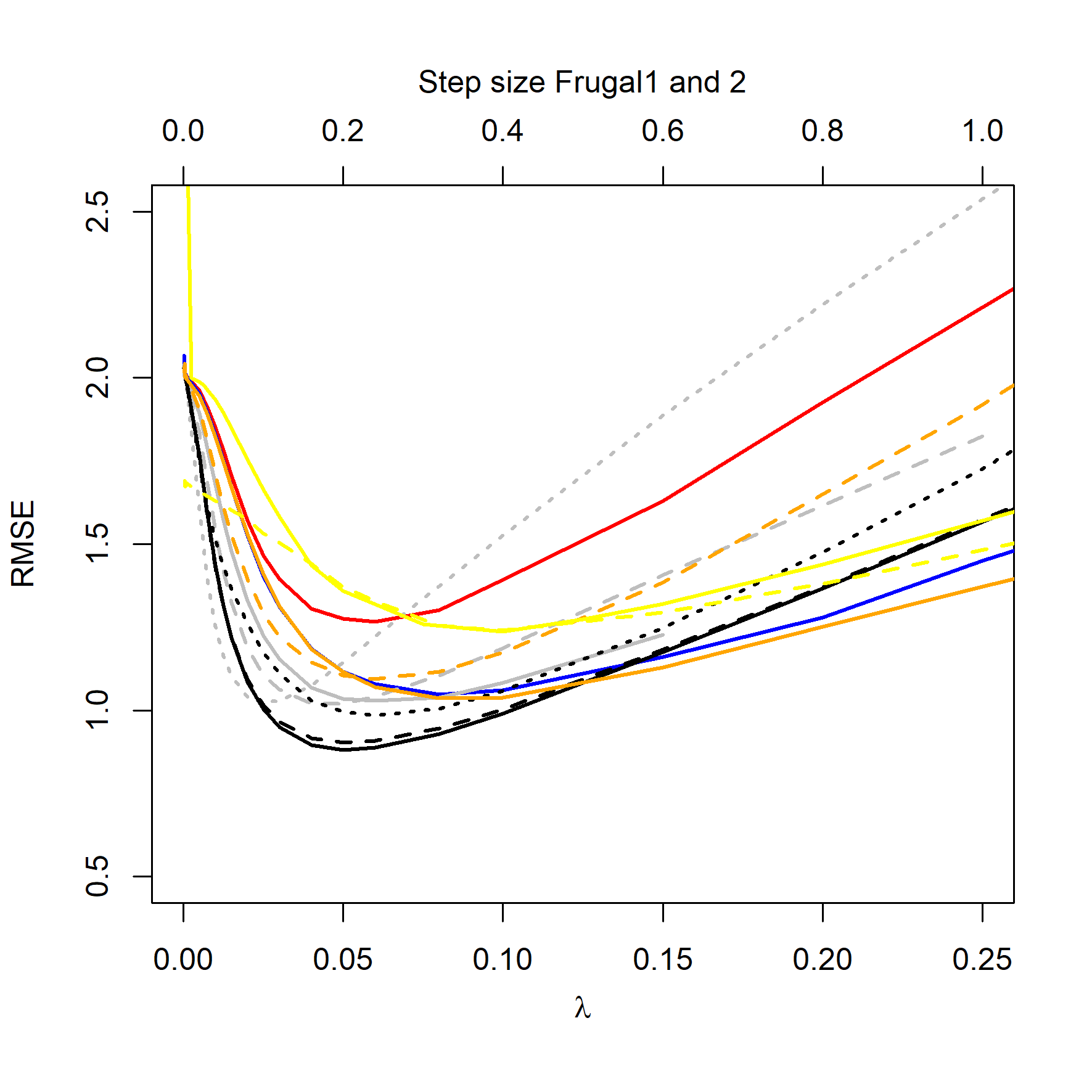} \\
   \includegraphics[width = 0.5\textwidth]{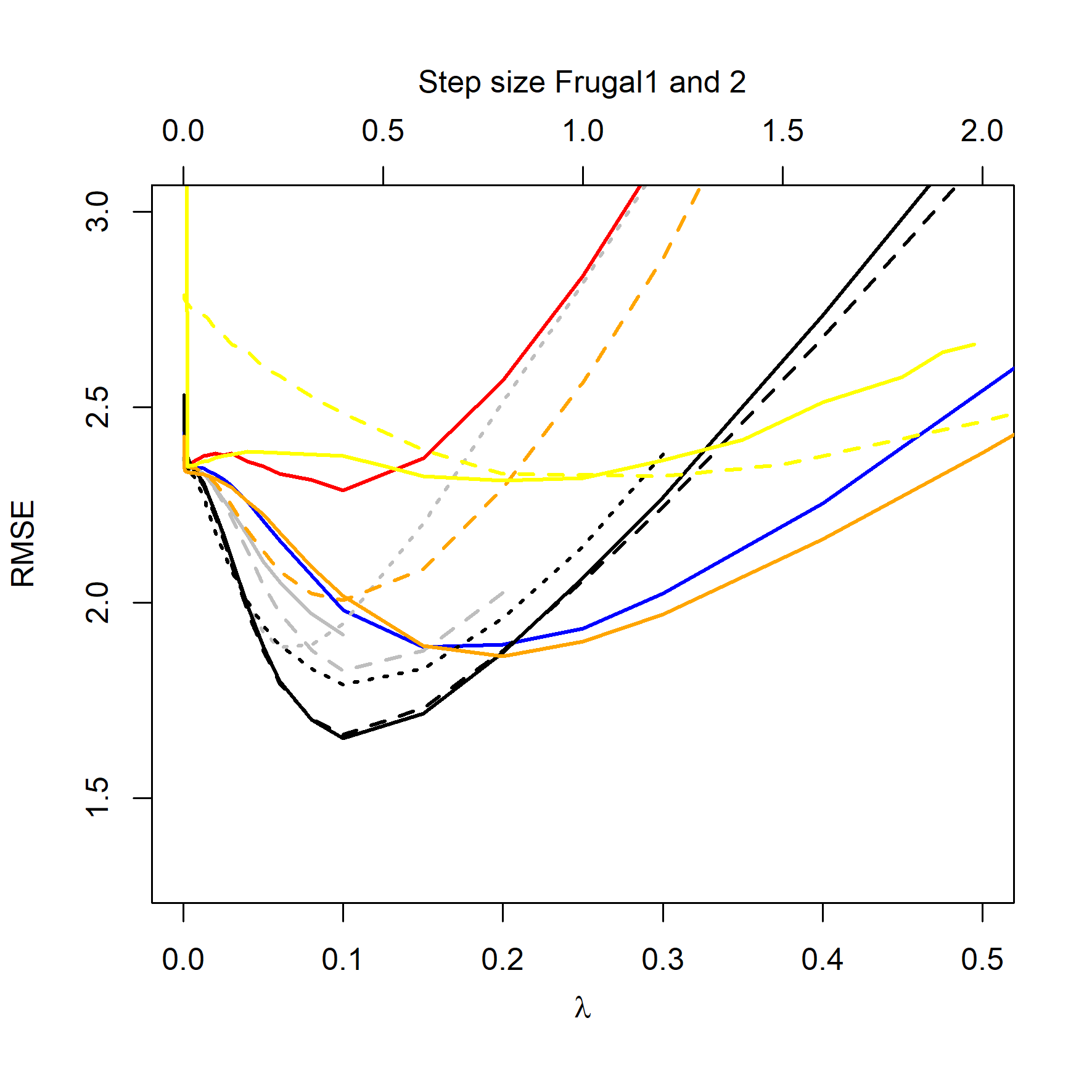} & \includegraphics[width = 0.5\textwidth]{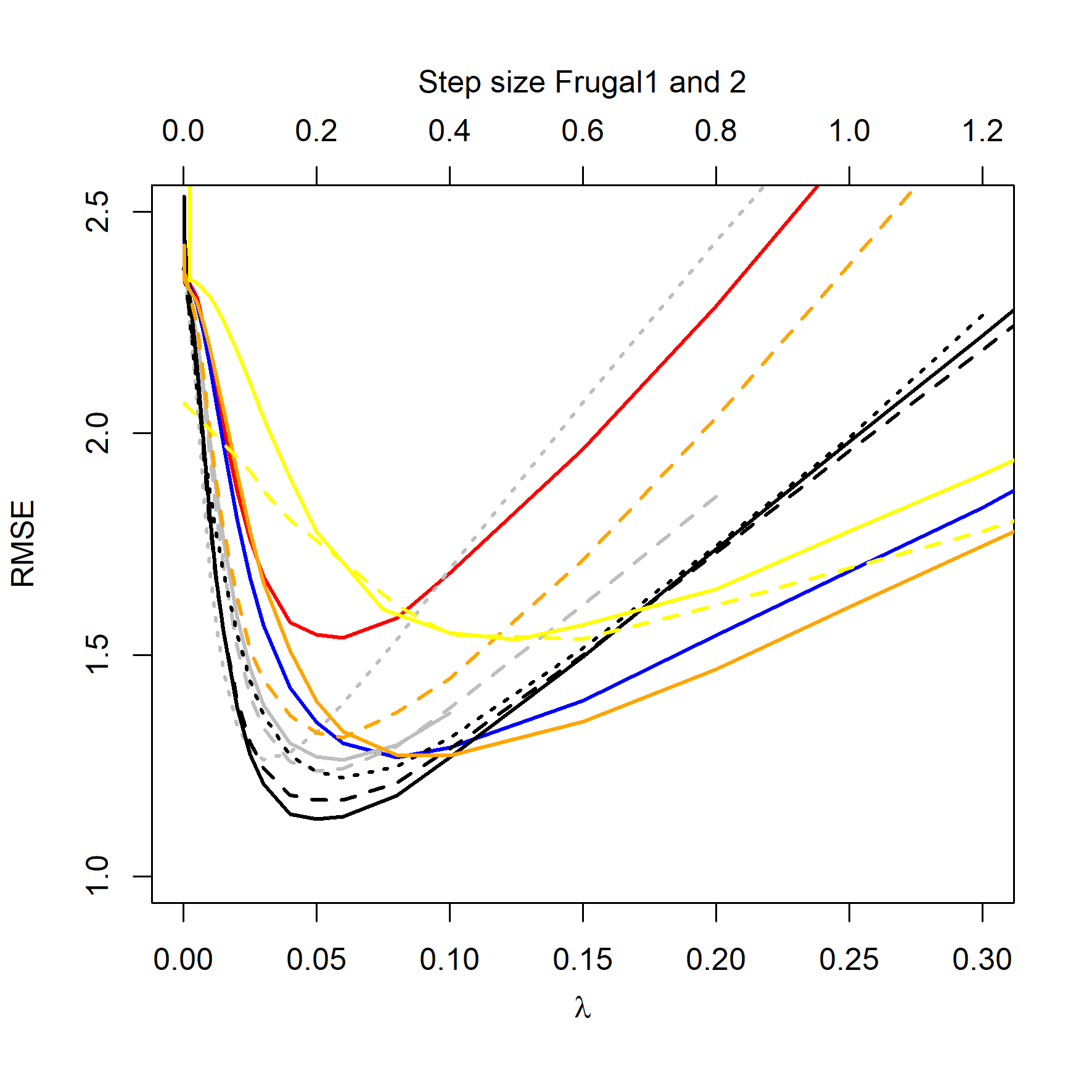} \\
   \includegraphics[width = 0.5\textwidth]{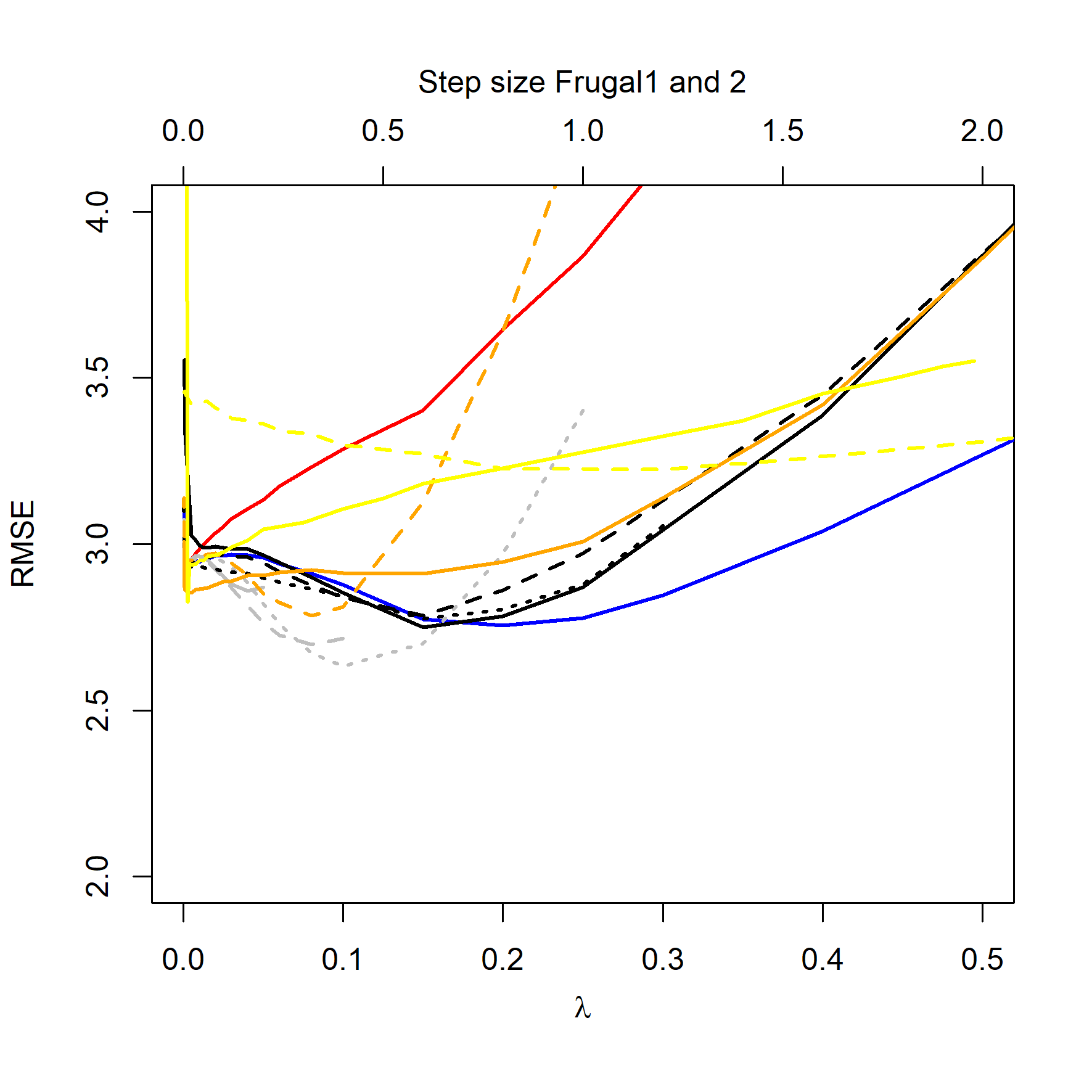} & \includegraphics[width = 0.5\textwidth]{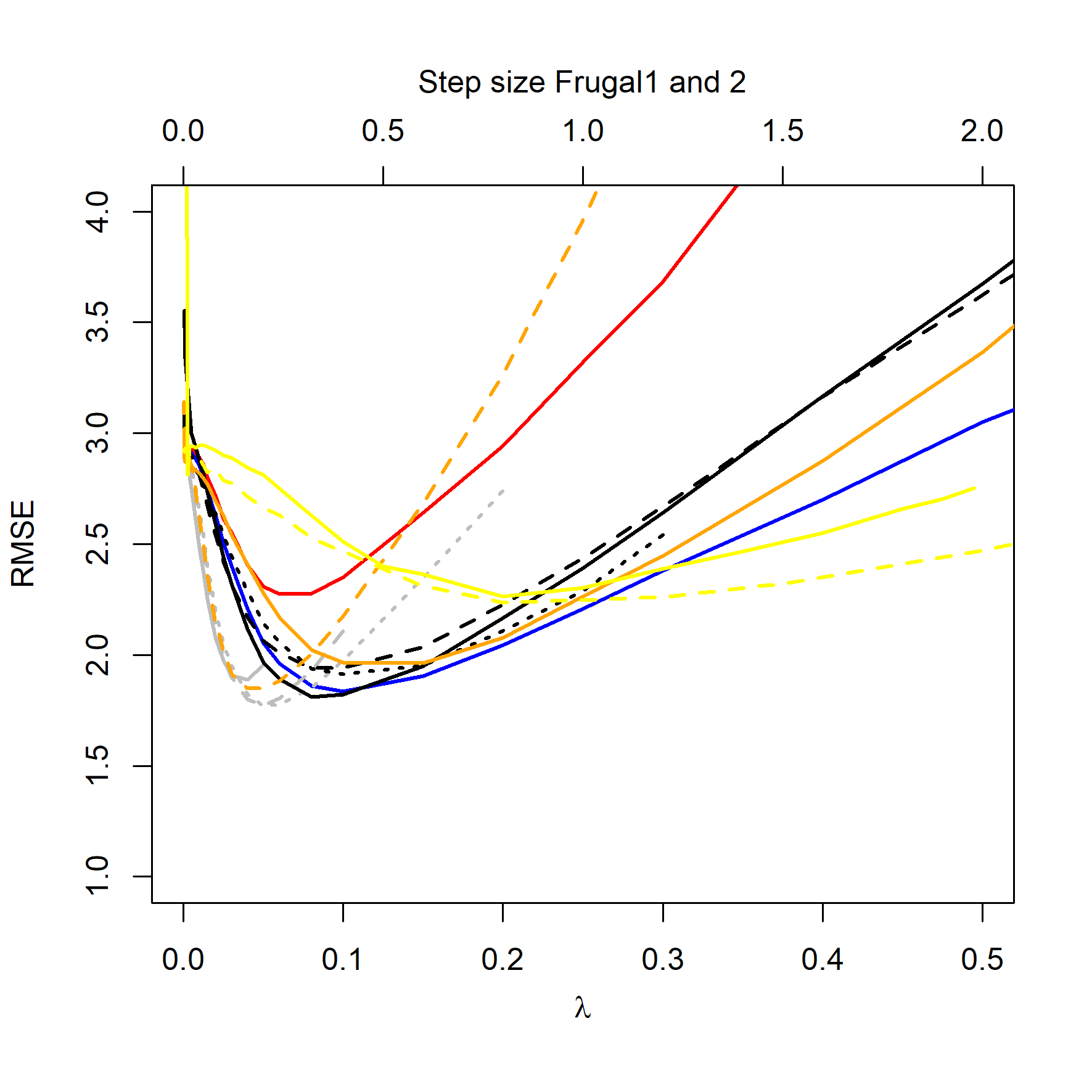}
  \end{tabular}
  \caption{$\chi^2$ distribution switch case: The left and right columns show results for $T=100$ and $T=500$, respectively. The rows from top to bottom show results when estimating quantile $Q_n(q=0.5),\,\, Q_n(q=0.7)$ and $Q_n(q=0.9)$, respectively. Ratio refers to the ratio between the tuning parameters, i.e. ratio = $\gamma/\lambda$. The upper $x$ axis refers to the step size in the Frugal algorithms.}
  \label{fig:5}
\end{figure}
Figures \ref{fig:2} to \ref{fig:5} illustrate the results of our experiments. For the normal distribution period case (Figure \ref{fig:2}), we see that the QEWA algorithm outperforms all the algorithms in the literature.
In accordance with the analysis in Section \ref{sec:qea}, the QEWA algorithm performed the best using a small value of the ratio $\gamma/\lambda$. The Cao et al. algorithm struggled with numerical problems for some choices of the tuning parameters and therefore some of the curves are short. 

For the normal distribution switch case (Figure \ref{fig:3}), we see that the QEWA algorithm again outperforms all the algorithms in the literature. Again we see that the QEWA performs best using a small value of the ratio $\gamma/\lambda$. 

For the $\chi^2$ distribution cases we see that the QEWA algorithm also here outperforms the other algorithms. For $q=0.9$, the QEWA algorithm documents competitive results to the best performing alternative algorithms. Also here a small value of the ratio $\gamma/\lambda$ is the preferable choice.

Among the alternative algorithms there are no consistency in which algorithm are closest to the performance of the QEWA, but overall the DUMIQE and DQTRE seem to be closes. However, all the alternative algorithms suffer with significantly poorer results than the QEWA for at least some cases. E.g. DQTRE performs poorly when estimating quanties in the tails ($q = 0.9$) and DUMIQE for the switch cases. 

\begin{table}[h]
    \centering
    \begin{tabular}{cccc}
        & $q = 0.5$  & $q = 0.7$ & $q = 0.9$  \\ \hline
      $T=100$ &   1.4278 & 1.5279 & 1.7646\\
      $T=500$ &   1.4233 & 1.5433 & 1.7342\\
    \end{tabular}
    \caption{Normal distribution periodic case: Root mean squared estimation error for the selection algorithm \cite{guha2009stream}.}
    \label{tab:2}
\end{table}
\begin{table}[h]
    \centering
    \begin{tabular}{cccc}
        & $q = 0.5$  & $q = 0.7$ & $q = 0.9$  \\ \hline
      $T=100$ & 2.0541 & 2.3171 & 2.5479 \\
      $T=500$ & 2.0947 & 2.3489 & 2.5427 \\
    \end{tabular}
    \caption{Normal distribution switch case: Root mean squared estimation error for the selection algorithm \cite{guha2009stream}.}
    \label{tab:3}
\end{table}
\begin{table}[h]
    \centering
    \begin{tabular}{cccc}
        & $q = 0.5$  & $q = 0.7$ & $q = 0.9$  \\ \hline
      $T=100$ & 1.4441 & 1.7423 & 2.4316\\
      $T=500$ & 1.4386 & 1.7273 & 2.6951\\
    \end{tabular}
    \caption{$\chi^2$ distribution periodic case: Root mean squared estimation error for the selection algorithm \cite{guha2009stream}.}
    \label{tab:4}
\end{table}
\begin{table}[h]
    \centering
    \begin{tabular}{cccc}
        & $q = 0.5$  & $q = 0.7$ & $q = 0.9$  \\ \hline
      $T=100$ & 2.0367 & 2.3913 & 3.3717\\
      $T=500$ & 2.0462 & 2.4137 & 3.1166\\
    \end{tabular}
    \caption{$\chi^2$ distribution switch case: Root mean squared estimation error for the selection algorithm \cite{guha2009stream}.}
    \label{tab:5}
\end{table}
Tables \ref{tab:2} to \ref{tab:5} show results for the selection algorithm \cite{guha2009stream}. The algorithm does not have any tuning parameters and the results thus are presented in tables. We see that QEWA outperforms the selection algorithm with a clear margin for all the different cases.

In summary the QEWA algorithm outperforms all the different state-of-the-art algorithms from the literature. Best performance is achieved using a small value of the ratio $\gamma/\lambda$.

\section{Real-life Data Experiments -- Concept Drift Detection}
\label{sec:real-life}

In most challenging data prediction tasks, the relation between input and output data evolves over time. Thus if static relationships are assumed, prediction performance will degrade with time. In the field of machine learning and data mining this phenomenon is referred to as concept drift \cite{gama2014survey}. Different strategies have been suggested to detect when the performance of the predictive model degrades and thus should be retrained/updated \cite{gama2014survey}. Current state-of-the-art strategies monitor the \textit{average} predictive error, but for real-life applications it is often more relevant to control that the prediction error rarely \textit{goes above} some critical threshold. In this example we demonstrate how to perform concept drift detection and adaptation on such a critical threshold by tracking an upper quantile of the prediction error distribution, e.g. the 80\% quantile. As an application domain, we investigate the case of efficient control of indoor climate.

Heating, ventilation and air conditioning (HVAC) systems typically control indoor climate by reacting on the current room conditions such as indoor temperature. However, given the time required for a HVAC system to adjust to changes in the indoor climate, such strategies always will lag behind resulting in poor control of indoor climate and energy usage. This raises the need for building models that forecast future indoor climate temperature and use this as input to the HVAC system. Zamora-Mart{\'\i}nez et al. \cite{zamora2014line} propose to use artificial neural network (ANN) models to forecast future indoor temperature based on a total of 20 features including outdoor climate variables such as temperature and precipitation amounts and indoor climates variables such as CO$_2$ level. Since more observations are received with time and the relation between input and output may evolve with time, the model is retrained in an online manner. The authors however do not take advantage of concept drift detection in order to efficiently decide when to retrain the model.

We now demonstrate how the suggested quantile estimator in this paper can be used for concept drift detection for the online indoor temperature forecasting problem described above. We consider the same dataset as in \cite{zamora2014line} where new observation of input and output variables is received every 15 minutes. We forecasted indoor temperature 15 minutes into the future using an autoregressive (AR) model of order one. In addition to the current indoor temperature, the current value of the other 20 features were used as input to the forecasting model. Given the large number of features, regularization of the model parameters was required to get a reliable forecasts and we relied on LASSO regularization \cite{friedman2010regularization}\footnote[3]{This model is a simple and natural forecasting model, but other and more advanced machine learning models that predict on the continuous scale, like ANN models, could also be used.}.

\begin{figure}
  \centering
  \begin{tabular}{cc}
    \includegraphics[width = 0.5\textwidth]{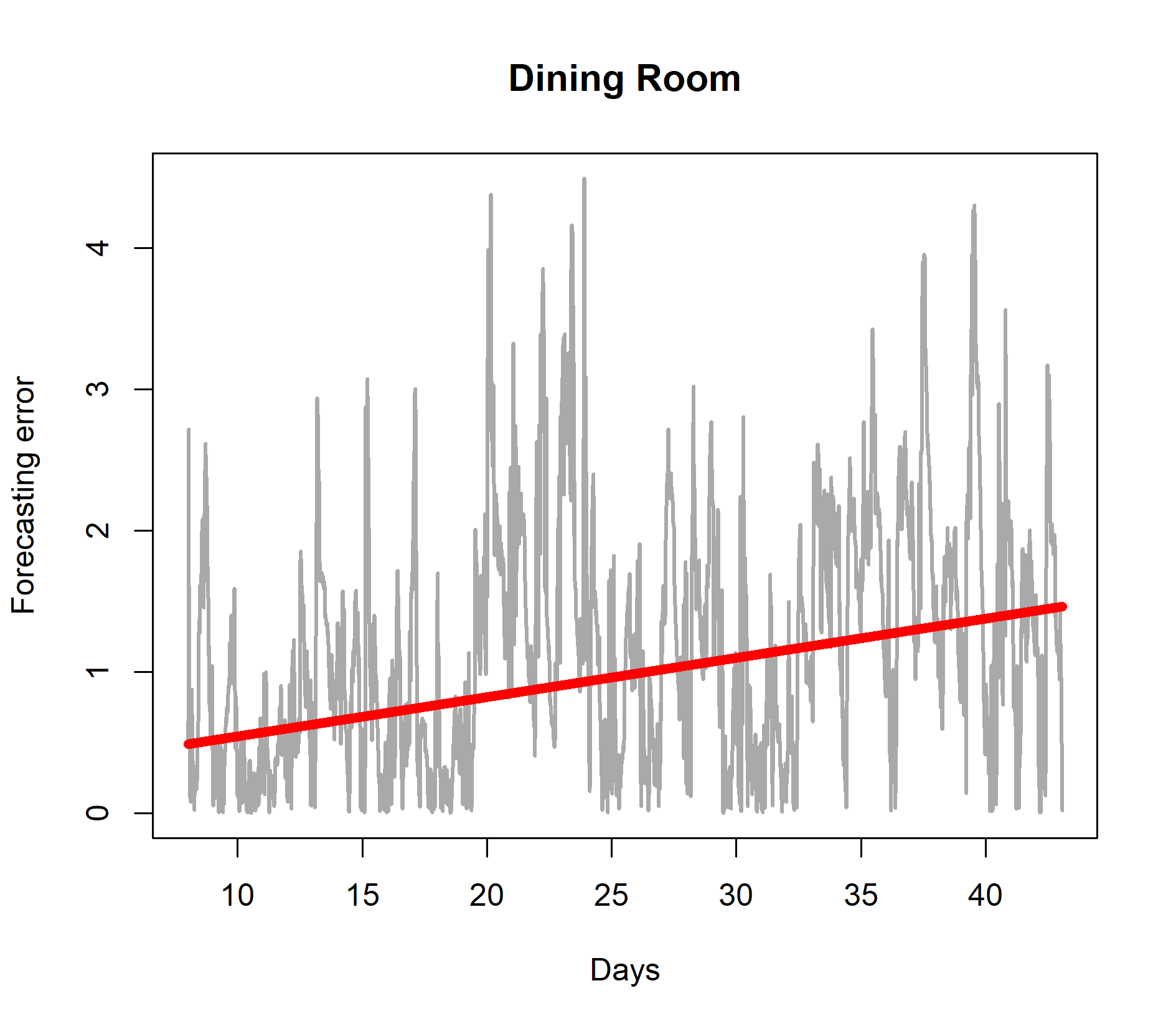} & \includegraphics[width = 0.5\textwidth]{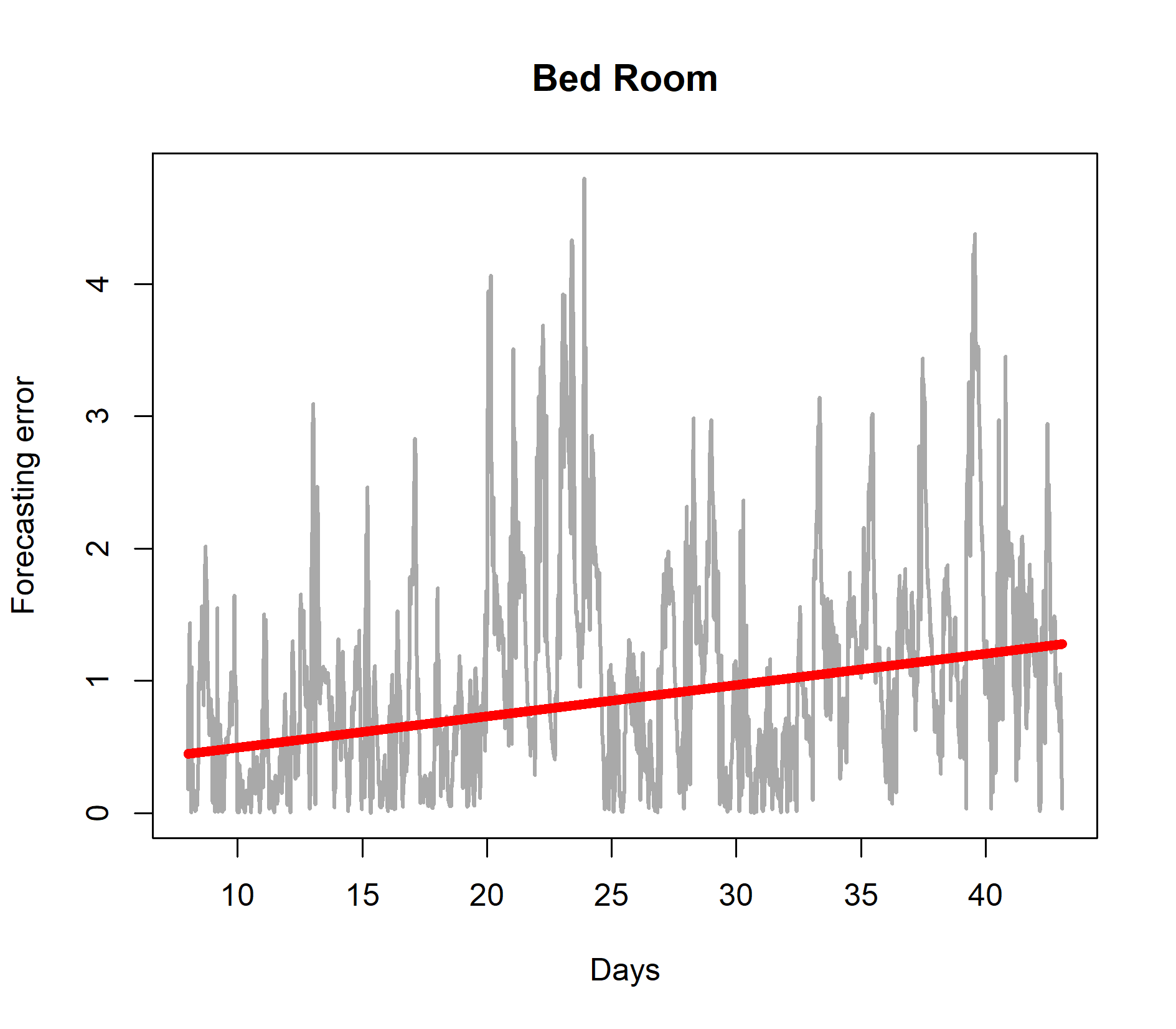}
  \end{tabular}
  \caption{The left and right panels refer to dining room and bed room, respectively. The x-axis refers to the number of days since the observation started. The gray curves show the forecasting error predicting 15 minutes into the future. The red curves show the linear trends in the forecasting error.}
  \label{fig:6}
\end{figure}
First, we trained the LASSO AR model based on eight days of observations and used the model to predict 15 minutes into the future each time a new observation was received. The results are shown in Figure \ref{fig:6}. The figure demonstrates that if the model is not retrained after day eight, the forecasting error gradually increases with time (the red line). In other words, the data is subject to concept drift and the forecasting model should be retrained as more observations are received. Instead of retraining the model regularly according to a fixed periodicity which is clearly ineffective, a sophisticated approach consists of retraining the model only if concept drift is detected.

\begin{figure}
  \centering
  \begin{tabular}{cc}
    \includegraphics[width = 0.5\textwidth]{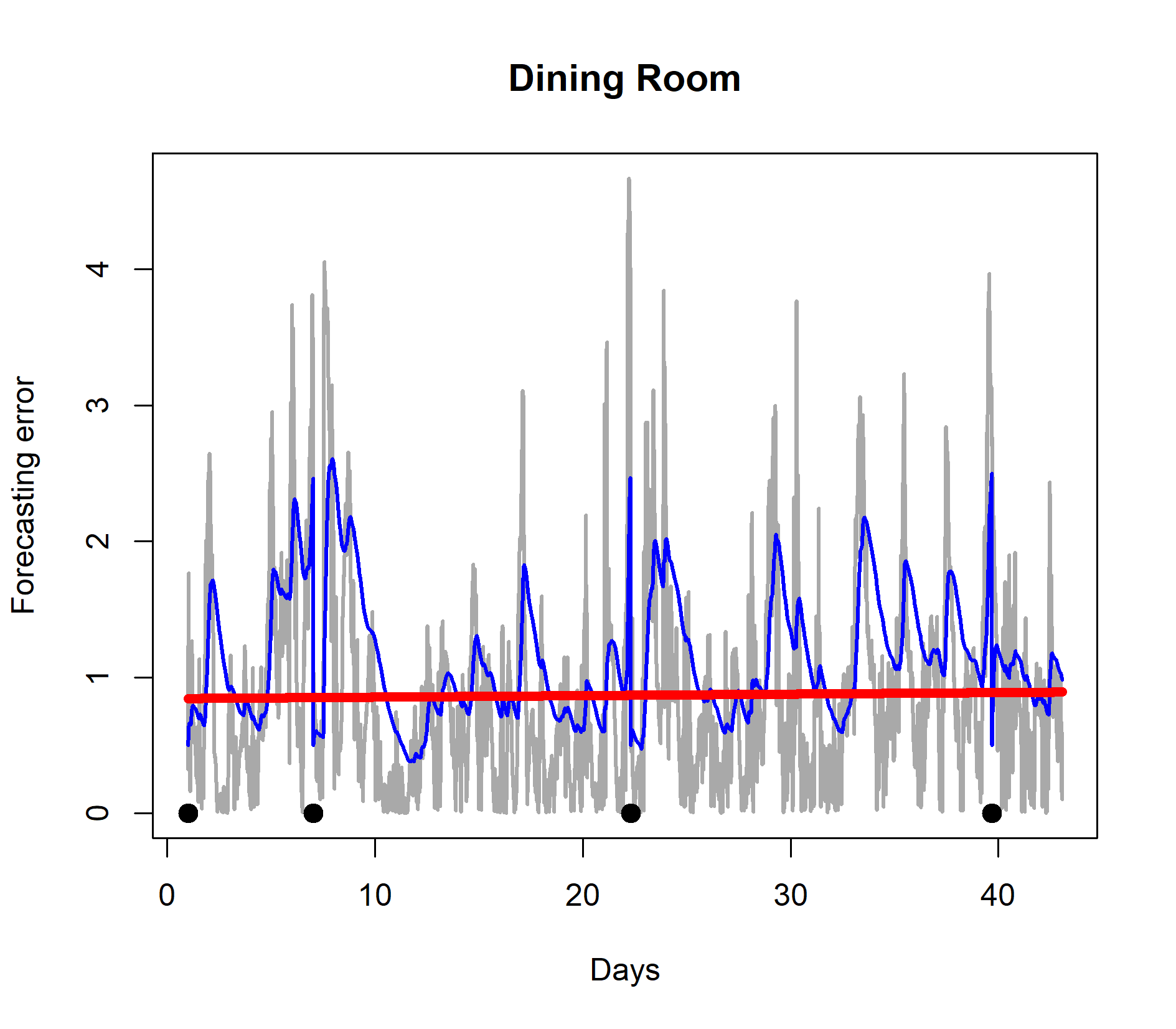} & \includegraphics[width = 0.5\textwidth]{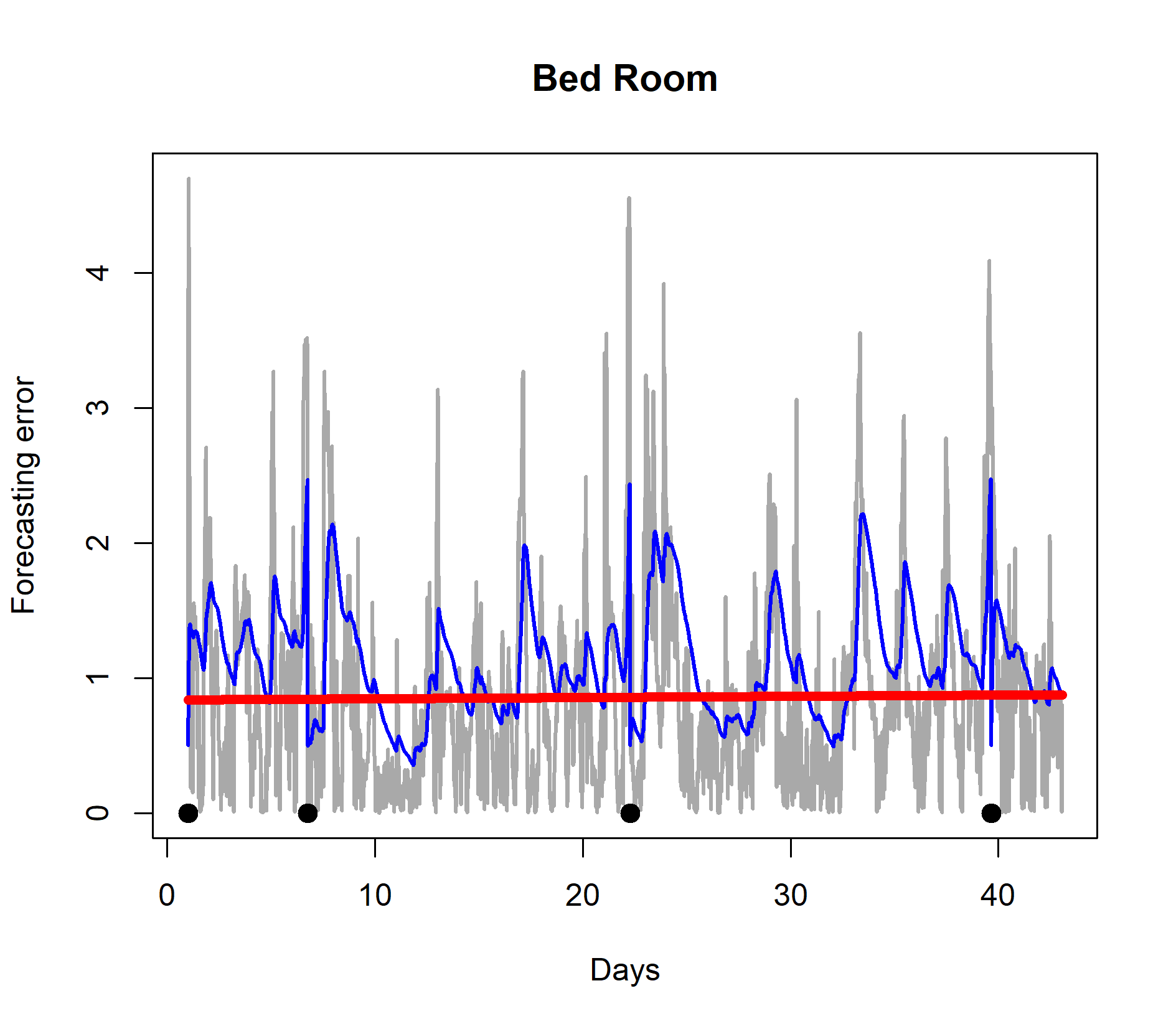}
  \end{tabular}
  \caption{The left and right panels refer to dining room and bed room, respectively. The x-axis refers to the number of days since the observation started. The gray curves show the forecasting error predicting 15 minutes into the future. The blue curves show tracking of the 80\% quantiles of the forecasting error data streams. The black dots along the x-axis show when the model was retrained. The red curves show the linear trends in the forecasting error.}
  \label{fig:7}
\end{figure}

We now build a concept drift and model retraining procedure based on quantile tracking. We required that the indoor temperature forecasting error rarely should go above two degrees centigrade. We used the QEWA estimator to track the 80\% quantile of the forecasting error data stream (the gray curves in Figure \ref{fig:6}). If the quantile estimate went above two degrees centigrade, the model was retrained. We trained the model for the first time after 24 hours of observations. The results are shown in Figure \ref{fig:7}. After the initial training after 24 hours of observations, the $80\%$ quantile estimate of the forecasting error distribution went above two degrees three times and each time the model was retrained. The results demonstrates that by a few selected retrainings of the model, the forecasting error is controlled, indicated by a horizontal linear trend (red curves) in Figure \ref{fig:7}.

In conclusion, the example demonstrates how the suggested quantile estimator can be useful for concept drift detection and model adaptation.

\section{Closing remarks}

The exponentially weighted moving average of observations is known to be the state-of-art estimator to track the expectation of dynamically varying data stream distributions. In this paper, we have presented an incremental quantile estimator that is in fact a generalized exponential weighted moving average estimator. To the best of our knowledge, this is the first quantile estimator in the literature that falls within this well-known class of efficient estimators. The experiments show that the estimator outperforms  state-of-the-art quantile estimators in the literature.

We demonstrate how tracking of quantiles has application in the field of machine learning. More particularly, we show how the suggested estimator can be used for tracking quantiles of the prediction error distribution in order to detect when a machine learning model should be retrained.

A potential ally for future research is to extend the QEWA estimator to simultaneously track multiple quantiles. One could of course, just run the QEWA estimator for each quantile of interest, but this could potentially lead to a violation of the monotone property of quantiles. The monotone property of quantiles, refers to the requirement that an estimate of a higher quantile should be always bigger than an estimate of a lower quantile e.g. the 50\% quantile always be above the 30\% quantile.

\clearpage

\bibliographystyle{plain}
\bibliography{bibl}

\clearpage

\appendix
\section{Proof of Theorem \ref{thm:1}}
\label{app:proof}

We will first present a theorem due to Norman \cite{Norman1972} that will be used to prove Theorem \ref{thm:1}.
Norman \cite{Norman1972} studied distance "diminishing models". The convergence of $\widehat{Q}_n(q)$ to $Q(q)$ is a consequence of this theorem.
\begin{theorem}
\label{thm:Norman}
Let $x(t)$  be a stationary Markov process dependent on a constant parameter $\theta \in [0,1]$. Each $x(t) \in I$, where $I$ is  a  subset  of  the  real  line.  Let $\delta x(t)=x(t+1)-x(t)$. The following are assumed to hold:
\begin{enumerate}
\item I is compact
\item $E [\delta x(t) | x(t)=y]= \theta w(y)+ O(\theta^2)$
\item $Var [\delta x(t) | x(t)=y]= \theta ^2 s(y)+ O(\theta^2)$
\item $E [\delta x(t)^3 | x(t)=y]=  O(\theta ^3)$
where $sup_{y \in I} \frac{O(\theta^k)}{\theta^k}< \infty$ for $k=2,3$ and $sup_{y \in I} \frac{o(\theta^2)}{\theta^2} \rightarrow 0$ as $\theta \rightarrow 0$
\item $w(y)$ has a Lipschitz derivative in $I$
\item $s(y)$ is Lipschitz $I$.
\end{enumerate}
If Assumptions 1 to 6 above hold, $w(y)$ has a unique root $y^*$ in $I$ and
$\frac{d w}{d y}  \bigg|_{y=y^*} \le 0$ then
\begin{enumerate}
\item $var [\delta x(t) | x(0)=x]=O(\theta)$ uniformly for all $x \in I$ and $t \ge 0$.
For any $x \in I$,  the differential equation $\frac{d y(\tau)}{d \tau}=w(y(t))$ has a unique solution  $y(\tau)=y(\tau,x)$ with $y(0)=x$  and	$E [\delta x(t) | x(0)=x]=y( t \theta)+O(\theta)$ uniformly for all $x \in I$ and  $t \ge 0$.
\item $\frac{x(t)-y(t \theta)}{\sqrt \theta}$ has a normal distribution with zero mean and finite variance as $\theta \rightarrow 0$ and $t \theta \rightarrow \infty$.
\end{enumerate}
\end{theorem}
Having presented Theorem \ref{thm:Norman}, we are now ready to prove Theorem \ref{thm:1}.
\begin{proof}
We now start by showing that the Markov process based on the updating rules in Eq. \eqref{eq:2} and Theorem \ref{thm:1} satisfies the assumptions 1 to 6 in Theorem \ref{thm:Norman}. We start by verifying assumption 2
\begin{align}
\notag
  &E\left(\delta \widehat{Q}_n(q)\,\left|\,\widehat{Q}_n(q)\right.\right) = \\
\notag
  & = E\left(\delta \widehat{Q}_n(q)\,\left|\,\widehat{Q}_n(q) \geq X\right. \right) P\left(\widehat{Q}_n(q) \geq X \right) + E\left(\delta \widehat{Q}_n(q)\,|\,\widehat{Q}_n(q) < X\right) P\left(\widehat{Q}_n(q) < X\right) = \\[2mm]
  \notag
  & = \lambda c_n\hspace{-0.5mm}\left(\widehat{Q}_{n}(q)\right) \frac{q}{\mu_n^+ - \widehat{Q}_{n}(q)} \left(\mu_n^+ - \widehat{Q}_{n}(q) \right)\left(1 - F\left(\widehat{Q}_n(q)\right)\right)\\[2mm]
\notag
  & - \lambda c_n\hspace{-0.5mm}\left(\widehat{Q}_{n}(q)\right)\frac{1-q}{\widehat{Q}_{n}(q) - \mu_n^-}   \left(\widehat{Q}_{n}(q) - \mu_n^- \right) F\left(\widehat{Q}_n(q)\right)\\[2mm]
\label{eq:10}
  &= \lambda c_n\hspace{-0.5mm}\left(\widehat{Q}_{n}(q)\right) \left(q - F\left(\widehat{Q}_n(q)\right)\right)
\end{align}
where $c_n\hspace{-0.5mm}\left(\widehat{Q}_{n}(q)\right)$ is as given in Eq. \eqref{eq:30}. We now let $\theta = \lambda$, $y = \widehat{Q}_n(q)$ and $w\hspace{-0.5mm}\left(\widehat{Q}_n(q)\right)$ equal to "everything" in Eq. \eqref{eq:10} except $\lambda$. It is easy to see that assumption 2 in Theorem \ref{thm:Norman} is satisfied. Further, since $\mu_n^+ - \widehat{Q}_{n}(q) > 0$ and $\widehat{Q}_{n}(q) - \mu_n^- > 0$, $w\hspace{-0.5mm}\left(\widehat{Q}_n(q)\right)$ has a
Lipschitz derivative and assumption 5 is satisfied.

Next we turn to assumption 3.
\begin{align}
  &\notag E\left(\delta \widehat{Q}_n(q)^2\,\left|\,\widehat{Q}_n(q)\right.\right) = \\
  &\notag  = E\left(\delta \widehat{Q}_n(q)^2\,\left|\,\widehat{Q}_n(q) \geq X \right.\right) P\left(\widehat{Q}_n(q) \geq X \right) + E\left(\delta \widehat{Q}_n(q)^2\,\left|\,\widehat{Q}_n(q) < X\right.\right) P\left(\widehat{Q}_n(q) < X\right) = \\
  \begin{split}
    \label{eq:16}
  & = \lambda^2 \left(c_n\hspace{-0.5mm}\left(\widehat{Q}_{n}(q)\right) \frac{q}{\mu_n^+ - \widehat{Q}_n(q)}\right)^2 \left(\mu_n^{2,+} - 2\widehat{Q}_n(q)\mu_n^+ + \widehat{Q}_n(q)^2 \right) \left(1 - F\left(\widehat{Q}_n(q)\right)\right) \\
  & + \lambda^2 \left(c_n\hspace{-0.5mm}\left(\widehat{Q}_{n}(q)\right) \frac{q}{\widehat{Q}_n(q) - \mu_n^-}\right)^2 \left(\widehat{Q}_n(q)^2 - 2\widehat{Q}_n(q)\mu_n^- + \mu_n^{2,-}\right) F\left(\widehat{Q}_n(q)\right)
  \end{split}
\end{align}
where $\mu_n^{2,+} = E(X^2_n|X_n > \widehat{Q}_{n}(q))$ and $\mu_n^{2,-} = E(X^2_n|X_n < \widehat{Q}_{n}(q))$. Further we know that
\begin{align}
  \label{eq:13}
  \begin{split}
  & Var\left(\delta \widehat{Q}{n} (q)\,\left|\,\widehat{Q}_n(q)\right.\right) = E\left(\delta \widehat{Q}{n} (q)^2\,\left|\,\widehat{Q}n (q)\right.\right) - E\left(\delta \widehat{Q}_{n} (q)\,\left|\,\widehat{Q}_n(q)\right.\right)^2
  \end{split}
\end{align}
By substituting Eq. \eqref{eq:10} and Eq. \eqref{eq:16} into Eq. \eqref{eq:13}, we see that assumption 3 is satisfied with $s\hspace{-0.7mm}\left(\widehat{Q}_{n}(q)\right)$ equal to everything in Eq. \eqref{eq:13} except $\lambda^2$. Since $\mu_n^+ - \widehat{Q}_{n}(q) > 0$ and $\widehat{Q}_{n}(q) - \mu_n^- > 0$, $s\hspace{-0.7mm}\left(\widehat{Q}_{n}(q)\right)$ is Lipschitz and assumption 6 is also satisfied. Assumption 4 can now be proved in the same manner.

We will use the results of Norman to prove the convergence. It is easy to see that $w\hspace{-0.5mm}\left(\widehat{Q}_{n} (q)\right)$ in Eq. \eqref{eq:10} admits one unique root $\widehat{Q}_n (q) = {F}^{-1}(q) = Q(q)$ $\left(\text{note } c_n\hspace{-0.5mm}\left(\widehat{Q}_{n}(q)\right) > 0\,\, \forall \widehat{Q}_{n}(q)\right)$.

We now differentiate to get:
\begin{align*}
  \frac{d\, w\left(\widehat{Q}_n(q)\right)}{d\, \widehat{Q}_n(q) } &= c'_n\hspace{-0.5mm}\left(\widehat{Q}_{n}(q)\right) \left(q - F\left(\widehat{Q}_n(q)\right)\right) - c_n\hspace{-0.5mm}\left(\widehat{Q}_{n}(q)\right)f\left(\widehat{Q}_n(q)\right)
\end{align*}
We substitute the unique root $Q(q)$ for $\widehat{Q}_n(q)$ and get
\begin{align*}
  \frac{ d\, w\left(\widehat{Q}_{n}(q)\right) }{d\, \widehat{Q}_{n}(q)} \bigg|_{\widehat{Q}_{n} (q)=Q(q)} &= c'_n\hspace{-0.5mm}\left(Q_{n}(q)\right) \left(q - F\left(Q_{n}(q)\right)\right) - c_n\hspace{-0.5mm}\left(Q_{n}(q)\right)f\left(Q_{n}(q)\right)\\
  & = 0 - c_n\hspace{-0.5mm}\left(Q_{n}(q)\right)f\left(Q_{n}(q)\right) < 0
\end{align*}
This gives
\begin{align*}
  \lim_{n \lambda  \to \infty, \lambda \to 0}  E\left(\widehat{Q}_n(q)\right)=Q(q)+O(\lambda)
\end{align*}
and
\begin{align*}
Var\left(\widehat{Q}_n(q)\right)=O(\lambda)
\end{align*}
Consequently
\begin{align*}
  \lim_{n \lambda  \to \infty, \lambda \to 0}  \widehat{Q}_n(q)=Q(q)
\end{align*}
\end{proof}

\end{document}